\documentclass[11pt,a4]{article}
\usepackage{amsmath}
\usepackage{amsgen,amstext,amsbsy,amsopn,amsthm, amssymb}

\numberwithin{equation}{section}
\newcommand{\version}{June 14, 2006 }
\newcommand{\R}{{\mathord{\mathbb R}}}

\newcommand{\be}{\begin{equation}}
\newcommand{\ee}{\end{equation}}
\newcommand{\bea}{\begin{eqnarray}}
\newcommand{\eea}{\end{eqnarray}}
\newcommand{\atd}{a_{2\rm D}}
\newcommand{\x}{{\bf x}}
\newcommand{\xp}{x}

\newcommand{\X}{{ X}}
\newcommand{\half}{\mbox{$\frac{1}{2}$}}
\newcommand{\beqa}{\begin{eqnarray}}
\newcommand{\eeqa}{\end{eqnarray}}

\newcommand{\beq}{\begin{equation}}
\newcommand{\eeq}{\end{equation}}

\newcommand{\eps}{\varepsilon}
\newcommand{\suli}{\sum\limits}

\newcommand{\xij}{|x_i-x_j|}
\newcommand{\V}{V}
\def\o1{{\mathrm{o}(1)}}

\newtheorem{thm}{Theorem}[section]
\newtheorem{lem}[thm]{Lemma}

\newtheorem{corollary}[thm]{Corollary}

\begin{document}
     %\markboth{\scriptsize{SY \version}}{\scriptsize{SY \version}}

\title{Bosons in Disc-Shaped Traps: From 3D to 2D}
\author{\hspace{-.2 cm} K. Schnee${}^{a}$, J. Yngvason${}^{a, b}$\\
\normalsize\it \hspace{-.5 cm}\\
\hspace{-.5 cm}\normalsize\it ${}^{a}$ Erwin Schr{\"o}dinger
Institute for Mathematical Physics,\\ \normalsize\it Boltzmanngasse 9,
1090 Vienna, Austria\\ ${}^{b}$\normalsize\it Institut f\"ur Theoretische
Physik, Universit{\"a}t Wien,\\ \normalsize\it Boltzmanngasse 5,
1090 Vienna, Austria}

\date{\version}

\maketitle
\vspace{0.4cm}
\begin{abstract}
We present a mathematically rigorous analysis of the ground state
of a dilute, interacting  Bose gas in a three-dimensional trap
that is strongly confining in one direction so that the system
becomes effectively two-dimensional.  The parameters involved are
the particle number, $N\gg 1$, the two-dimensional extension,
$\bar L$, of the gas cloud in the trap, the thickness, $h\ll \bar
L$ of the trap, and the scattering length $a$ of the interaction
potential. Our analysis starts from the full many-body Hamiltonian
with an interaction potential that is assumed to be repulsive,
radially symmetric and of short range, but otherwise arbitrary. In
particular, hard cores are allowed. Under the premises that the
confining energy, $\sim 1/h^2$, is much larger than the internal
energy per particle, and $a/h\to 0$, we prove that the system can
be treated as a gas of two-dimensional bosons with scattering
length $a_{\rm 2D}= h\exp(-(\hbox{\rm const.)}h/a)$. In the parameter region
where $a/h\ll |\ln(\bar\rho h^2)|^{-1}$, with $\bar\rho\sim N/\bar
L^2$ the mean density, the system is described by a
two-dimensional Gross-Pitaevskii density functional with coupling
parameter $\sim Na/h$.  If $|\ln(\bar\rho h^2)|^{-1}\lesssim a/h$
the coupling parameter is $\sim N |\ln(\bar\rho h^2)|^{-1}$ and
thus independent of $a$.  In both cases Bose-Einstein condensation
in the ground state holds, provided the coupling parameter stays
bounded.
\end{abstract}
\input epsf
\section{Introduction}\label{sect:1}
In recent experiments dilute Bose gases have been confined in
magneto-optical traps in such a way that the particle motion is
essentially frozen in one or two directions and the system becomes
effectively lower dimensional \cite{goerlitz, greiner, schreck,
esslinger, paredes, kinoshita, naegerl}. This is an intrinsically
quantum mechanical phenomenon because it is not necessary to have
a trap width or thickness that is the size of an atom, but it
suffices that the energy gap for the motion in the strongly
confined direction(s) is large compared to the internal energy per
particle. The case of highly elongated (``cigar-shaped'') traps
has received particular attention \cite{olshanii, dunjko, Das1,
das2, girardeau2, petrov, LSY} (see also \cite{LSY} for further
references), because it opens the possibility to realize the
one-dimensional Lieb-Liniger model \cite{LL}, and even the
limiting Girardeau-Tonks case \cite{gir, kinoshita, paredes} that
exhibits strong correlations.  A detailed, rigorous derivation of
the one-dimensional behavior from the many-body Hamiltonian of a
three-dimensional gas was given in the paper \cite{LSY}.  This is
not a simple problem, one reason being that an approximate
factorization of the ground state wave function in the
longitudinal and transverse variables is in general not possible
(in particular not for hard core potentials), and the proofs in
\cite{LSY} are, in fact, quite long.

In the present paper we carry out a corresponding analysis for
thin, disc-shaped traps, i.e., traps with strong confinement in
one direction so that a two-dimensional behavior is expected.
Experimental realizations of such systems and possible mechanisms
for creating them are discussed, e.g., in \cite{goerlitz, greiner,
colombe, ZG, naegerl}.  On the theoretical side the references
\cite{Petrov, olshaniipricup, PiSt2003, pricoup, PGS} contain many
valuable insights into their properties.  There are several
similarities with the emergence of one-dimensional behavior in
cigar-shaped traps, but also some notable differences.  Like for
cigar-shaped traps, there is a basic division of the parameter
domain into two regions: one where a limit of a three-dimensional
Gross-Pitaevskii (GP) theory applies, and a complementary region
described by a ``truly'' low dimensional theory. In the case
discussed in \cite{LSY} the latter is a density functional theory
based on the exact Lieb-Liniger solution for the energy of a
strongly interacting (and highly correlated) one-dimensional gas
with delta interactions. (Note that in 1D strong interactions
means low density.) In the present case, on the other hand, the
gas is {\it weakly} interacting in all parameter regions.  In the
region not accessible from 3D GP theory the energy formula
\cite{schick, LY2d, LSY2d} for a dilute two-dimensional Bose gas
with a logarithmic dependence on the density applies.  To enter
this region extreme dilution is required.  The Lieb-Liniger region
in the 1D case demands also quite dilute systems, but the
requirement is even more stringent in 2D. This will be explained
further below.

  We recall from \cite{schick, LY2d} that the energy per particle of a dilute,
  homogeneous, two-dimensional Bose gas with density $\rho_{\rm 2D}$
  and scattering length $a_{\rm 2D}$ of the interaction potential is
  (in units such that $\hbar=2m=1$) \be\label{SCHNEE2den} e_{\rm
  2D}\approx 4\pi\rho_{\rm 2D}|\ln(\rho_{\rm 2D}a_{\rm 2D}^2)|^{-1}.
  \ee The corresponding result in three dimensions is {
  \be\label{SCHNEE3den} e_{\rm 3D}\approx 4\pi\rho_{\rm 3D}a_{\rm 3D},
  \ee cf.\ \cite{LY1998}.  In the following we shall denote the
  two-dimensional density, $\rho_{\rm 2D}$, simply by $\rho$ and the
  three dimensional scattering length, $a_{\rm 3D}$, by $a$.

  Our results establish rigorously, and in the many-body context, a
  relation between $a$, the thickness, $h$, of the trap (assumed to be
  $\gg a$) and the effective two-dimensional scattering length,
  $a_{\rm 2D}$.  Essentially, as $h$ tends to zero, $a_{\rm 2D}=
  h\exp(-{(\rm const.)}h/a)$.  (The
  precise formula is given in Eq.\ \eqref{SCHNEEatd} below.)  If $|\ln
  \rho h^2|\ll h/a$, then $|\ln(\rho a_{\rm 2D}^2)|\approx h/a$, and
  the two-dimensional formula \eqref{SCHNEE2den} leads to the same
  result as the three dimensional formula \eqref{SCHNEE3den}, because
  $\rho_{\rm 3D}\sim \rho/h$.  The ``true'' two dimensional region
  requires $|\ln \rho h^2|\gtrsim h/a$ and hence the condition
  $\rho^{-1/2}\gtrsim h e^{h/a}$ for the interparticle distance,
  $\rho^{-1/2}$.  This should be compared with the corresponding
  condition for the 1D Lieb-Liniger region in \cite{LSY}
  where the interparticle distance is ``only'' required to be of the
  order or larger than $h^2/a$.

  The basic formula $a_{\rm 2D}= h\exp(-({\rm const.})h/a)$ for the
  scattering length appeared, to the best of our knowledge, first in
  \cite{Petrov}. It can be motivated by considering a weak, bounded
  potential, where perturbation theory can be used to compute the
  energy for a two-body problem that is directly related to the
  scattering length, cf. Appendix A in \cite{LY2d}. This perturbative
  calculation is carried out in Section \ref{pertscatt} as a step in
  the proof of a lower bound for the many-body energy; its relation to
  the formula for $a_{\rm 2D}$ is explained in the Remark after
  Corollary \ref{SCHNEEscattcorr}. We wish to stress, however, that
  deriving this formula in the context of two-body scattering is only
  a step towards the solution of the many-body problem that is the
  concern of the present paper.

We now define the setting and state the results more precisely.  We
consider $N$ identical, spinless bosons in a confining,
three-dimensional trap potential and with a repulsive, rotationally
symmetric pair interaction.  We take the direction of strong
confinement as the $z$-direction and write the points $\x\in\R^3$ as
$(\xp,z)$, $\xp\in\R^2$, $z\in\R$.  The Hamiltonian is
\begin{equation}\label{SCHNEEdef:ham}
H_{N,L,h,a} = \sum_{i=1}^N \left(-\Delta_i + V_{L,h}(\x_i) \right)
+ \sum_{1\le
i <j \le N} v_a(|\x_i-\x_j|)
\end{equation}
with
\begin{eqnarray}\label{SCHNEEpot}
V_{L,h}(\x) &=& V_L(\xp)+V^\perp_h(z) =
\frac{1}{L^2}
V(L^{-1}\xp)+ \frac{1}{h^2}V^{\perp
}(h^{-1}z),\\ \label{va}
v_a(|\x|) &=& \frac{1}{a^2}v(a^{-1}|\x|).
\end{eqnarray}
The confining potentials $V$ and $V^\perp$ are
assumed to be locally
bounded and tend to $\infty$ as $|\xp|$ and $|z|$
tend to $\infty$.
The interaction potential $v$ is assumed to be nonnegative, of finite range and
with
scattering length 1; the scaled potential
$v_a$ then has scattering length $a$. We regard $v$,
$V^{\perp}$ and $V$ as fixed and $L,h,a$ as scaling
parameters.  The Hamiltonian (\ref{SCHNEEdef:ham}) acts on symmetric
wave functions in $L^2(\R^{3N},d\x_1\cdots d\x_N)$. Its
ground state energy,  $E^{\rm QM}({N,L,h,a})$, scales with $L$ as
\be\label{SCHNEEqmscal} E^{\rm QM}(N,L,h,a)=\frac1{L^2} E^{\rm QM}(N,1,h/L,a/L).\ee
\\
Taking $N\to\infty$ but keeping $h/L$ and $Na/L$ fixed leads to a
three dimensional Gross-Pitaevskii description of the ground state as
proved in \cite{LSY1999}.  The corresponding energy functional is
\be\label{SCHNEE3dgp}{\mathcal E}^{\rm GP}_{\rm
3D}[\phi]=\int_{\R^3}\left\{|\nabla\phi(\x)|^2+V_{L,h}(\x)|\phi(\x)|^2+4\pi
Na |\phi(\x)|^4\right\}d^3\x\ee and the energy per particle is
\beqa \nonumber
E^{\rm GP}_{\rm 3D}(N,L,h,a)/N &=&\inf\{{\mathcal
E}^{\rm GP}_{\rm 3D}[\phi]:\,\, \hbox{$\int|\phi(\x)|^2
d^2\x$}=1\}\\ \label{SCHNEE3dgpen}&=&(1/L^2) E^{\rm GP}_{\rm 3D}(1,1,h/L,Na/L).\eeqa
By Theorem 1.1 in \cite{LSY1999}, we have, for fixed $h/L$ and $Na/L$,
\be\label{SCHNEE3dgplim}\lim_{N\to\infty}\frac {E^{\rm
QM}({N,L,h,a})}{E^{\rm GP}_{\rm 3D}(N,L,h,a)}=1.  \ee It is important
to note, however, that the estimates in \cite{LSY1999} are not uniform in
the ratio $h/L$ and the question what happens if $h/L\to 0$ is not
addressed in that paper.  It will be shown in the next section that a
{\it part} of the parameter range for thin traps can be treated by
considering, at fixed $Na/h$, the $h/L\to 0$ limit of $E^{\rm GP}_{\rm
3D}(1,1,h/L,Na/L)$, with the ground state energy for the transverse
motion, $\sim 1/h^2$, subtracted.  But this limit can evidently never
lead to a logarithmic dependence on the density and it does not give
the correct limit formula for the energy in the whole parameter range.

To cover all cases we have to consider a two-dimensional
Gross-Pitaevskii theory of the type studied in \cite{LSY2d}, i.e.,
\be\label{SCHNEE2dgp}{\mathcal E}^{\rm GP}_{\rm
2D}[\varphi]=\int_{\R^2}\left\{|\nabla\varphi(x)|^2+V_{L}(x)|\varphi(x)|^2+4\pi
Ng |\varphi(x)|^4\right\}d^2x\ee with \be\label{SCHNEEcoupl}
g=|\ln(\bar\rho\atd^2|^{-1}.\ee\\
Here $\bar\rho$ is the mean density, defined as in Eq.\ (1.6) in
\cite{LSY2d}. An explicit formula that is valid in the case $Ng\gg 1$ can be states as follows. Let
\be \rho^{\rm TF}_N(x)=\frac1{8\pi}\left[\mu^{\rm TF}_N-V_L(x)\right]_+ \ee
be the 'Thomas Fermi' density for $N$ particles at coupling constant 1 
in the potential $V_L$, where $\mu^{\rm TF}_N$ is chosen so that
$\int \rho^{\rm TF}_N=N$. Then
\be\label{barrho1} \bar\rho=N^{-1}\int_{\R^2} \rho^{\rm TF}_N(x)^2 dx.\ee
%It is essentially
%the same as $N\int\varphi^{\rm GP}(x)^4 dx$
%where $\varphi^{\rm GP}$ is the minimizer of
%\eqref{SCHNEE2dgp} with $\int\varphi^2=1$. (Cf.\ the remarks
%following Eq.\ (1.6) in \cite{LSY2d}.)
For simplicity we shall assume that $V$ is homogeneous
of some degree $p>0$, i.e., $V(\lambda x)=\lambda^p V(x)$, and in this
case \be\label{SCHNEErhobar}\bar\rho\sim N^{p/(p+2)}/L^2 =N/\bar
L^2\quad\hbox{with}\quad \bar L=N^{1/(p+2)}L.\ee The length $\bar L$
measures the effective extension of the gas cloud of the $N$ particles in the
two-dimensional trap.  A box potential corresponds to $L=\bar L$,
i.e., $p=\infty$ and hence $\bar\rho\sim N/L^2$. 

The case  $Ng=O(1)$ requires a closer look at the definition of $\bar\rho$. First, for any value of  $Ng$ we can consider the minimizer $\varphi^{\rm GP}_{Ng}$ of  \eqref{SCHNEE2dgp} with normalization $\Vert \varphi^{\rm GP}_{Ng}\Vert_2=1$. The corresponding mean density is
\be \bar\rho_{Ng}=N\int |\varphi^{\rm GP}_{Ng}|^4.\ee
A general definition of $\bar \rho$ amounts to solving the equation $\bar\rho=\bar\rho_{Ng}$ with $g$ as in \eqref{SCHNEEcoupl}. As discussed in \cite{LSY2d} this gives the same result as \eqref{barrho1} to leading order in $g$ when $Ng\gg 1$. In the case $|\ln Nh/L|\ll h/a$ (referred to as 'Region I' below) the coupling constant is simply $a/h$ and thus independent of $N$. Moreover, in  a homogeneous potential of degree $p$ the effective length scale $\bar L$ is $\sim (Ng+1)^{1/(p+2)} L$ and thus of  order $L$ if $Ng=O(1)$.

 The energy per
particle corresponding to \eqref{SCHNEE2dgp} is \be\label{SCHNEE2dgpen} E^{\rm
GP}_{\rm 2D}(N,L,g)/N=\inf\{{\mathcal E}^{\rm GP}_{\rm
2D}[\varphi]:\,\, \hbox{$\int|\varphi(x)|^2 d^2x$}=1\}=(1/L^2) E^{\rm
GP}_{\rm 2D}(1,1,Ng).\ee

Let $s_{h}$ be the normalized ground state wave function of the one-particle
Hamiltonian $-d^2/dz^2+V^\perp_h(z)$. It can be written as
$s_{h}(z)=h^{-1/2}s(h^{-1}z)$ and the ground state energy as
$e^\perp_{h}=h^{-2}e^\perp$, where $s(z)$ and $e^\perp$ are,
respectively, the ground state wave function and ground state energy of
$-d^2/dz^2+V^\perp(z)$. We {\it define} the two dimensional
scattering length by the formula
\be \label{SCHNEEatd}\atd=h\exp\left(-(\hbox{$\int s(z)^4dz$})^{-1}h/2a\right).\ee
Then, using (\ref{SCHNEEcoupl}), \be\label{SCHNEEgformula}
g=|-\ln(\bar\rho h^2)+\hbox{($\int s(z)^4dz$})^{-1}h/a|^{-1}.\ee
The justification of the definition (\ref{SCHNEEatd}) is
Theorem \ref{SCHNEEthm:main} below.\\

\noindent{\it Remark.} Since $\atd$ appears only under a logarithm,
and $a/h\to 0$, one could, at least as far as leading order
computations are concerned, equally well define the two dimensional
scattering length as $\atd'=b\,\exp\left(-(\hbox{$\int
s(z)^4dz$})^{-1}h/2a\right)$ with $b$ satisfying $c\, a\leq b\leq C\,
h$ for some constants $c>0$, $C<\infty$.  In fact, if
$g'=|\ln(\bar\rho(\atd')^2)|^{-1}$, then \be\label{ggprime} \frac
{g}{g'}=1+\frac{2\ln(b/h)} {|-\ln(\bar \rho
h^2)+\hbox{(const.)}h/a|}\to 1\ee because $(a/h)\ln(b/h)\to 0$.

\medskip

We can now state the limit theorem for the ground state energy:
\begin{thm}[From 3D to 2D, ground state energy]\label{SCHNEEthm:main}
Let $N\to\infty$ and at the same time $h/L\to 0$ and $a/h\to 0$ in
such a way that $h^2\bar \rho g\to 0$ (with $g$ given by Eq.\
\eqref{SCHNEEgformula}). Then
\begin{equation}\label{SCHNEElimit}
\lim  \frac{E^{\rm QM}(N,L,h,a) - N h^{-2} e^{\perp}}{E^{\rm GP}_{\rm
2D}(N,L,g)} =1.
\end{equation}
\end{thm}

\noindent{\it Remarks:} 1.  The condition $h^2\bar \rho g\to 0$
means that the ground state energy $h^{-2}e^\perp$ associated with
the confining potential in the $z$-direction is much larger than
the energy $\bar \rho g$.  This is the condition of {\it strong
confinement} \index{strong confinement} in the $z$-direction.  In
the case that $h/a\gg |\ln(\bar\rho h^2)|$ we have $g\sim a/h$ and
hence the condition in that region is equivalent to \be
\label{SCHNEEahll}\bar\rho ah\ll 1.\ee On the other hand, if
$h/a\lesssim |\ln(\bar\rho h^2)|$ the strong confinement condition
is equivalent to $h^2\bar\rho|\ln( h^2\bar\rho)|^{-1}\ll 1$, which
means simply that \be\label{SCHNEEhhll} \bar\rho h^2\ll 1\,. \ee
Both \eqref{SCHNEEahll} and \eqref{SCHNEEhhll} clearly imply $\bar
\rho\atd^2\ll 1$, i.e., the gas is dilute in the 2D sense (and
also in the 3D sense, $\rho_{\rm 3D}a^3\ll 1$, because $\rho_{\rm
3D}=\rho/h$).  This is different from the situation in
cigar-shaped traps considered in \cite{LSY} where the gas can be
either dilute or dense in the 1D sense, depending on the
parameters (although it is always dilute in the 3D sense).

\medskip
\noindent 2.  It is, in fact, not necessary to demand $h/L\to 0$
explicitly in Theorem \ref{SCHNEEthm:main}.  The reason is as follows.
In the region where $h/a\lesssim |\ln(\bar\rho h^2)|$, the strong
confinement condition $\bar\rho h^2\ll 1$ immediately implies $h/L\ll
1$ because $\bar\rho\gg 1/L^2$, cf.\ Eq.\ (\ref{SCHNEErhobar}).  If
$h/a\gg |\ln(\bar\rho h^2)|$, then at least $\bar\rho ah\ll 1$ holds
true.  This leaves only the alternatives $h/L\to 0$, or, if $h/L$
stays bounded away from zero, $Na/L\to 0$.  But the latter alternative
means, by the three dimensional Gross-Pitaevskii limit theorem
\cite{LSY1999}, that
the energy converges to the energy of a noninteracting, trapped gas,
for which \eqref{SCHNEElimit} obviously holds true.

\medskip

We shall refer to the parameter region where $h/a\gg |\ln(\bar\rho
h^2)|$ as {\bf Region I}, and the one where $h/a\lesssim |\ln(\bar\rho
h^2)|$ as {\bf Region II}.  In Region I we can take \be\label{SCHNEEgreg1}
g=(\hbox{$\int s(z)^4dz$})a/h.\ee In Region II $g\sim |\ln(\bar\rho
h^2)|^{-1}$, and in the extreme case that $h/a\ll |\ln(\bar\rho
h^2)|$, \be\label{SCHNEEgreg2} g=|\ln(\bar\rho h^2)|^{-1}.\ee In particular
$g$ is then independent of $a$ (but dependent on $\bar\rho$).  As
remarked earlier, Region II applies only to very dilute gases since it
requires interparticle distances $\bar\rho^{-1/2}\gtrsim h e^{h/a}$.

By Eq.\ \eqref{SCHNEE2dgpen} the relevant coupling parameter is $Ng$ rather
than $g$ itself, and both Region I and Region II can be divided
further, according to $Ng\ll 1$, $Ng\sim 1$, or $Ng\gg 1$.  The case
$Ng\ll 1$ corresponds simply to an ideal gas in the external trap
potential.  Note that this limit can both be reached from Region I by
taking $a/h\to 0$ at fixed $\bar\rho h^2$, or from Region II by
letting $\bar\rho h^2$ tend more rapidly to zero than $e^{-h/a}$.
The
case $Ng\sim 1$ in Region I corresponds to a GP theory with coupling
parameter $\sim Na/h$ as was already explained, in particular
after Eq.\ \eqref{SCHNEE3dgplim}.  The case $Ng\gg 1$ is the
`Thomas-Fermi'  case where the gradient term in the energy
functional \eqref{SCHNEE2dgp} can be ignored.  In Region II, the case
$Ng\lesssim 1$ requires $\bar\rho^{-1/2}\gtrsim h e^N$ and is thus
mainly of academic interest, while
$\bar\rho^{-1/2}\ll h e^N$ (but still $he^{h/a}\lesssim
\bar\rho^{-1/2}$) corresponds to the TF case.

The subdivision of the parameter range just described is somewhat different
from the situation described in \cite{LSY}. The reason is
the different form of the energy per particle of the low dimensional gas as
function of the density.\\

Theorem \ref{SCHNEEthm:main} is a limit theorem for the ground state
energy.  By standard arguments (variation with respect
to the trapping potential) it implies a convergence
result for the one particle
density in the ground state, $\rho^{\rm QM}_{N,L,h,a}(x)$ (cf.\ Section
2 in \cite{LSY} and
Theorem 8.2 in \cite{LSSY}):

Define the 2D QM density by integrating over the transverse variable
$z$, i.e., \beq \hat\rho^{\rm QM}_{N,L,h,a}(x):= \int \rho^{\rm
QM}_{N,L,h,a} (x,z)dz \ .  \eeq With $\bar L$ the extension of the
system in the 2D trap, cf.\  \eqref{SCHNEErhobar}, define the rescaled GP density
$\tilde\rho$ by \beq \tilde \rho(x)={\bar L^2}|\varphi^{\rm
GP}(\bar L x)|^2 \eeq
where $\varphi^{\rm GP}$ is the minimizer of \eqref{SCHNEE2dgp} with
normalization $\int |\varphi^{\rm GP}(x)|^2dx=1$.  (Note that $\tilde \rho$
depends on $N$, $L$ and $g$.)
\begin{thm}[{\bf 2D limit for the density}]\label{T2dens}
In the same limit as considered in Theorem~\ref{SCHNEEthm:main},
\beq
\lim \left( \frac {\bar L^2}N \hat\rho^{\rm QM}_{N,L,h,a}(\bar Lx)  -
\tilde \rho(x)\right) = 0
\eeq
in weak $L^1$ sense.
\end{thm}

In the GP case where the coupling parameter $Ng$ stays bounded a much
stronger result can be proved, namely convergence of the 1-particle
density matrix and Bose-Einstein condensation (BEC) in the ground
state. Recall that the one-body density matrix
obtained from the ground state wave function $\Psi_0$ is
\beq
\gamma_{\Psi_0}(\x,\x')=N\int_{\R^{3(N-1)}}
\Psi_0(\x,\x_2,\dots,\x_N)\Psi_0(\x',\x_2,\dots,\x_N)^*d\x_2\cdots
d\x_N.
\eeq
BEC means that in the $N\to \infty$ limit it
factorizes as $N\psi(\x)\psi(\x')$ for some normalized $\psi$. This,
in fact, is 100\% condensation and was proved in \cite{LS02} for a
fixed trap potential in the Gross-Pitaevskii limit, i.e., for both
$h/L$ and $Na/L$ fixed as $N\to\infty$. Here we extend this result to
the case $h/L\to 0$ with $Ng$ and $L$ fixed. The function $\psi$ is the
minimizer of the 2D GP functional (\ref{SCHNEE2dgp})
times the transverse function $s_h(z)$:

\begin{thm}[BEC in GP limit] \label{BEC}
If $N\to \infty$, $h/L\to 0$ while $Ng$ and $L$ are fixed, then
\beq\label{xx}
\frac{h}N\gamma_{\Psi_0}(x,hz;x',hz')\to
\varphi^{\rm GP}(x)\varphi^{\rm GP}(x')s(z)s(z')
\eeq
in trace norm. Here $\varphi^{\rm GP}$ is the normalized minimizer of the GP
functional (\ref{SCHNEE2dgp}).
\end{thm}

There is a variant of Theorem \ref{SCHNEEthm:main} that applies to
the thermodynamic limit in the 2D variable $x$ where the density
becomes
homogeneous in this variable. Let $E_{\rm
box}(N,L,a)$ be the ground state energy of the Hamiltonian
\eqref{SCHNEEdef:ham} with
the potential $V_{L}(x)$  replaced by a 2D box of side length $L$
and define, for fixed $\rho$, $h$ and $a$,
\beq e_{\rm{2D}}(\rho,h,a):=\lim_{N,L\to\infty, N/L^2=\rho}\frac{E_{\rm
box}(N,L,a)-Nh^{-2}e^\perp}{N}.\eeq
This is the energy per particle in the 2D thermodynamic limit
(with the confining energy subtracted) and
by standard arguments (cf.\ e.g., \cite{R}) it is independent of the boundary
conditions (Dirichlet, Neumann
or periodic)  imposed in the definition of $E_{\rm
box}(N,L,a)$. We then have

\begin{thm}[From 3D to 2D, homogeneous case]\label{SCHNEEthm:main2D}
    If $h^2\rho g\to 0$ and $a/h\to
0$, with $g$ given by Eq.\ \eqref{SCHNEEgformula}, then
\begin{equation}\label{SCHNEElimit2D}
\lim  \frac{e_{\rm 2D}(\rho,h,a)}{4\pi\rho g} =1.
\end{equation}
\end{thm}

The essentials of the dimensional reduction are already contained in
the proof of Theorem \ref{SCHNEEthm:main2D} and are more transparent
in this case than for inhomogeneous gases in Theorem
\ref{SCHNEEthm:main}.  Hence we shall focus on the proof of
Theorem \ref{SCHNEEthm:main2D}, while the modifications that have to
be made for a proof of Theorem \ref{SCHNEEthm:main} will be more
briefly described with appropriate references to \cite{LSY1999,
LSY2d, LSY, norway}, where analogous problems are discussed
in some detail.

An abbreviated version of this paper appears as Chapter 9 in
the Oberwolfach Seminars volume \cite{LSSY}.

\section{The 2D limit of 3D GP theory}

As in \cite{LSY}, certain aspects of the dimensional
reduction of the many-body system
can be seen already in the much simpler context of GP theory.
We therefore begin by considering the $h/L\to 0$ limit
of the 3D GP ground state energy. The result is, apart from the
confining energy, the 2D GP energy with coupling constant $g\sim a/h$.
This shows in particular that Region II, where $g\sim |\ln(\bar\rho
h^2)|^{-1}$, cannot be reached as a limit of 3D GP theory.

\begin{thm}[2D limit of 3D GP energy] Define $g=\left(\int s(z)^4 dz\right)a/h$. If
$h/L\to 0$, then
\be\frac{E_{\rm 3D}^{\rm
GP}(N,L,h,a)-Nh^{-2}e^\perp}{E_{\rm 2D}^{\rm
GP}(N,L,g)}\to 1\ee
uniformly in the parameters, as long as
$\bar\rho ah\to 0$.
\end{thm}

\noindent {\it Remark.} Since $E_{\rm 2D}^{\rm GP}(1,L,Ng)\sim L^{-2}+ \bar\rho a/h$,
 the condition $\bar\rho ah\to 0$ is equivalent to $h^2 E_{\rm
 2D}^{\rm GP}(1,L,Ng)\to 0$, which  means simply
 that the 2D GP energy per particle
 is much less than the confining energy, $\sim 1/h^2$.

\begin{proof}
Because of the scaling relation \eqref{SCHNEE3dgpen}
it suffices to consider the case $N=1$ and $L=1$.

For an upper bound to the 3D GP ground state energy we make the ansatz
\be\label{SCHNEEans}\phi(\x)= \varphi^{\rm GP}(x)s_{h}(z), \ee where
$\varphi^{\rm GP}$ is the minimizer of the 2D GP functional with
coupling constant $g$.  Then
\be{\mathcal E}^{\rm GP}_{\rm 3D}[\phi]=e^\perp/h^2+E_{\rm 2D}^{\rm
GP}(1,1,g)\ee and hence \be E_{\rm 3D}^{\rm GP}(1,1,h,a)-
e^\perp/h^2\leq E_{\rm 2D}^{\rm GP}(1,1,g).\ee

For the lower bound we consider the one-particle Hamiltonian (in 3D)
\be\label{SCHNEEhha} H_{h,a}=-\Delta+V_{1,h}(\x)+8\pi a|\varphi_{\rm
GP}(x)|^2 s_{h}(z)^2.  \ee Taking the 3D GP minimizer $\Phi$
as a test state gives
\bea\nonumber \hbox{inf spec}\, H_{h,a}&\leq& E_{\rm 3D}^{\rm
GP}(1,1,h,a)-4\pi a\int_{\R^3}|\Phi(\x)|^4 d^3\x\\&& +8\pi a\int |\varphi_{\rm
GP}(x)|^2s_{h}(z)^2|\Phi(\x)|^2d^3\x\nonumber
\\ &\leq& E_{\rm 3D}^{\rm
GP}(1,1,h,a)+4\pi a\int_{\R^3}|\varphi^{\rm GP}(x)|^4
s_{h}(z)^4d^3\x\nonumber \\ &=& E_{\rm 3D}^{\rm
GP}(1,1,h,a)+4\pi g\int_{\R^2}|\varphi^{\rm GP}(x)|^4 dx.\label{SCHNEEinfspec} \eea To
bound $H_{h,a}$ from below we consider first for fixed $x\in\R^2$ the
Hamiltonian (in 1D) \be H_{h,a,x}=-\partial_{z}^2+V_{h}^\perp(z)+8\pi a
|\varphi^{\rm GP}(x)|^2 s_{h}(z)^2.  \ee We regard
$-\partial_{z}^2+V_{h}^\perp(z)$ as its ``free'' part and $8\pi a
|\varphi^{\rm GP}(x)|^2 s_{h}(z)^2$ as a perturbation.  Since the
perturbation is positive all eigenvalues of $H_{h,a,x}$ are at least
as large as those of $-\partial_{z}^2+V_{h}(z)$; in particular the
first excited eigenvalue is $\sim 1/h^2$.
The expectation value in the ground state $s_{h}$ of the free part is
\be\label{SCHNEEhx}\langle H_{h,a,x}\rangle=e^\perp/h^2+8\pi g |\varphi_{\rm
GP}(x)|^2.\ee Temple's inequality \cite{TE, RS} gives
\be\label{SCHNEEtemplegp} H_{h,a,x}\geq \left(e^\perp/h^2+8\pi g
|\varphi^{\rm GP}(x)|^2\right)\left(1-\frac{\langle(H_{h,a,x}-\langle
H_{h,a,x}\rangle)^2\rangle}{\langle H_{h,a,x}\rangle(\tilde
e^\perp-e^\perp)/h^2}\right)\,,\ee where $\tilde e^\perp/h^2$ is the
lowest eigenvalue above the ground state energy of
$-\partial_{z}^2+V_{h}^\perp(z)$.  Since \be H_{h,a,x}s_{h}=(e^\perp/h^2)
s_{h}+8\pi a |\varphi^{\rm GP}(x)|^2 s_{h}^3\ee we have
$(H_{h,a,x}-\langle H_{h,a,x}\rangle)s_{h}=8\pi|\varphi_{\rm
GP}(x)|^2(as_{h}^3-gs_{h})$ and hence, using $g=a\int
s_{h}^4=(a/h)\int s^4$, \begin{eqnarray} \langle(H_{h,a,x}-\langle
H_{h,a,x}\rangle)^2\rangle&=&(8\pi)^2|\varphi_{\rm
GP}(x)|^4\int\left(as_{h}(z)^3-gs_{h}(z)\right)^2dz\nonumber\\ &\leq&
(8\pi)^2\Vert \varphi^{\rm GP}\Vert_{\infty}^4(a/h)^2\left[\int
s^6-\left(\int s^4\right)^2\right]\nonumber\\&\leq & \hbox{const.}\,E_{\rm
2D}^{\rm GP}(1,1,g)^2\end{eqnarray} where we have used Lemma 2.1 in
\cite{LSY2d} to bound the term $g\Vert \varphi^{\rm GP}\Vert_{\infty}^2$ by
$\hbox{const.}\,E_{\rm 2D}^{\rm GP}(1,1,g)$.  We thus see from
\eqref{SCHNEEhx} and the assumption $h^2 E_{\rm 2D}^{\rm GP}(1,1,g)\to 0$
that the error term in the Temple inequality \eqref{SCHNEEtemplegp} is
$o(1)$.

Now $H_{h,a}=-\Delta_{x}+V(x)+H_{h,a,x}$, so from
\eqref{SCHNEEtemplegp} we
conclude that \be \label{SCHNEEtemplegp2}H_{h,a}\geq \left((e^\perp/h^2)-\Delta_{x}+V(x)+8\pi g
|\varphi^{\rm GP}(x)|^2\right)(1-o(1)).\ee On the other hand, the
lowest energy of $-\Delta_{x}+V(x)+8\pi g |\varphi^{\rm GP}(x)|^2$ is just
$E_{\rm 2D}^{\rm GP}(1,1,g)+4\pi g\int_{\R^2}|\varphi_{\rm
GP}(x)|^4dx$.  Combining \eqref{SCHNEEinfspec} and \eqref{SCHNEEtemplegp2} we
thus get \be E_{\rm 3D}^{\rm GP}(1,1,h,a)- e^\perp/h^2\geq E_{\rm
2D}^{\rm GP}(1,1,g)(1-o(1)).\ee
\end{proof}

\iffalse
\noindent{\it Remark.} This proof holds also for the Gross-Pitaevskii
functional for rotating gases, i.e., if a rotational term, $-\langle
\phi,{\vec \Omega}\cdot{\vec L}\phi\rangle$ is added to the
functional.  Here $\vec \Omega$ is the angular velocity, assumed to
point in the $z$-direction, and ${\vec L}$ the angular momentum
operator.  The minimizer $\varphi^{\rm GP}$ is in this case complex
valued in general and may not be unique \cite{rot2}.
\fi

\section{Upper bound}\label{sect9.3}

\subsection{Finite ${\mathord{\hbox{\boldmath
$n$}}}$  bounds}\label{finitenupp}

We now turn to the many-body problem, i.e., the proof of Theorems
\ref{SCHNEEthm:main} and \ref{SCHNEEthm:main2D}.
Like in \cite{LSY} the key lemmas are
energy bounds in  boxes with {\it finite} particle number.
The bounds for the total system are obtained by distributing the
particles optimally among the boxes.  We start with the upper bound
for the energy in a single box.

Consider the Hamiltonian \be \label{SCHNEEboxham}H =  \sum_{i=1}^n
\left(- \Delta_i + V_{h}^\perp(z_i) \right) + \sum_{1\le i <j \le n}
v_a(|\x_i-\x_j|)\ee in a region $\Lambda=\Lambda_{2}\times\R$ where
$\Lambda_{2}$ denotes a box of side length $\ell$ in the 2D $x$
variables.  For the upper bound on the ground state energy of
(\ref{SCHNEEboxham}) we impose {\it Dirichlet} boundary conditions on
the 2D Laplacian.  The goal is to prove, for a given 2D density $\rho$
and parameters $a$ and $h$, that for a suitable choice of $\ell$ and
corresponding
particle number $n=\rho \ell^2$ the
energy per particle, with the confining energy $e^\perp/h^2$ 
subtracted,  is bounded above by \be\label{boxubound} 4\pi \rho|\ln(\rho
a_{\rm 2D}^2)|^{-1}(1+o(1)) \ee where $a_{\rm 2D}$ is
given by Eq.\ (\ref{SCHNEEatd}).  Moreover, the Dirichlet localization
energy per particle, $\sim 1/\ell^2$, should be small compared to
(\ref{boxubound}).  The relative error, $o(1)$, in (\ref{boxubound})
tends to zero with the small parameters $a/h$ and $\rho a h$ (Region
I), or $a/h$ and $\rho h^2$ (Region II).

The choice of variational functions depends on the
parameter regions and we are first concerned with the Region II, i.e.,
the case $|\ln(\rho h^2)|\gtrsim h/a$.

\subsubsection{Upper bound in Region II}

Let $f_{0}(r)$ be the solution of the zero energy scattering equation
\be\label{SCHNEEscatteq} -\Delta f_{0}+\half v_{a}f_{0}=0,\ee
normalized so that $f_{0}(r)=(1-a/r)$ for $r\geq R_{0}$ with $R_0$ the range of $v_a$. Note that $R_0=({\rm const.}) a$ by the scaling \eqref{va}.  The function $f_0$ satisfies
$0\leq f_{0}(r)\leq 1$ and $0\leq f_{0}'(r)\leq \min\{1/r,a/r^2\}$. This is seen by writing $f_0(r)=u(r)/r$ and 
$f'_0(r)=u'(r)/r-u(r)/r^2$ with $u^{\prime\prime}(r)=\half v(r)u(r)\geq 0$. Since $u(r)=r-a$ for $r\geq R_0$ and $u(0)=0$, convexity implies $u(r)\geq\max\{0,r-a\}$ and $u'(r)\leq 1$. Hence $0\leq f_{0}'(r)\leq \min\{1/r,a/r^2\}$ and $0\leq f_0(r)\leq\lim_{r\to\infty}(1-a/r)=1$.

For $R>R_{0}$ we define
\beq \label{eff} f(r)=f_{0}(r)/(1-a/R)\quad{\rm for}\quad 0\leq r\leq R,\quad
{\rm and}\quad f(r)=1\quad {\rm for} \quad r>R.\eeq  We define a two-dimensional potential by
\be\label{SCHNEEwpot}W(x)=\frac{2\Vert s
\Vert_4^4}h\int_{\R}\left[f'(|\x|)^2+\half
v_{a}(|\x|)f(|\x|)^2\right]dz.\ee Clearly, $W(x)\geq 0$, and $W$ is
rotationally symmetric with $W(x)=0$ for $|x|\geq R$.  Moreover, by
partial integration, using \eqref{SCHNEEscatteq}, it follows that
$W\in L^1(\R^2)$ with \be\label{SCHNEEintw}\int_{\R^2}W(x)dx=\frac
{8\pi a\Vert s \Vert_4^4}h (1-a/R)^{-1}.\ee

Define, for $b>R$,
\be\label{varphi}\varphi(r)=\left\{ \begin{array}{lcr}
{\ln(R/\atd)}/
{\ln(b/\atd)}&{\rm if} & 0\leq r\leq R\\ \\ {\ln(r/\atd)}/
{\ln(b/\atd)}&{\rm if}& R\leq r\leq b\\ \\ 1 &{\rm if} & b\leq r\end{array}
\right.\ee
As
test function for the three dimensional Hamiltonian \eqref{SCHNEEboxham} we shall
take
\be\label{SCHNEEFG}\Psi(\x_{1},\ldots,\x_{n})=F(\x_{1},\ldots,\x_{n})G(\x_{1},\ldots,\x_{n})\ee
with \be F(\x_1,\dots,\x_n)=\prod_{i<j} f(|\x_i-\x_j|) \quad
\hbox{and}\quad
G(\x_1,\dots,\x_n)=\prod_{i<j} \varphi(|x_i-x_j|)\prod_{k=1}^n s_h(z_k).\ee
The parameters $R$, $b$ and also $\ell$ will eventually be chosen so that the
errors compared to the expected leading term in the energy are small.

As it stands, the function (\ref{SCHNEEFG}) does not satisfy
Dirichlet boundary conditions but this can be taken care of by
multiplying the function with additional
factors at energy cost $\sim 1/\ell^2$ per particle, that will turn
out to be small compared to the energy of (\ref{SCHNEEFG}).

Since  $f(|\x_i-\x_j|) \varphi(|x_i-x_j|)=1$ for $|x_i-x_j|\geq b$
and $s_{h}$ is normalized, the
norm of $\Psi$ can be estimated as in Eq.\ (3.9) in \cite{LSY}.
\be\label{SCHNEEnormbd}
\langle\Psi|\Psi\rangle \geq
\ell^{2n}\left[1-\frac{\pi n(n-1)}2\frac {b^2}{\ell^2}\right]\,.\ee

Next we consider the expectation value of $H$ with the wave function
$\Psi$. By partial integration we have, for every $j$,
\bea\label{SCHNEEupp1}
\!\!\!\!&&\!\!\!\! \int|\nabla_{j}(FG)|^2\nonumber \\ \!\!\!\!&&\!\!\!\!=\int G^2|\nabla_{j}F|^2-\int
F^2G\Delta_{j}G=\int G^2|\nabla_{j}F|^2-\int
F^2G\,\partial_{z_{j}}^2G-\int F^2 G(\Delta^\parallel_{j} G)\nonumber \\ \!\!\!\!&&\!\!\!\!
=\int G^2|\nabla_{j}F|^2-\int
F^2G\,\partial_{z_{j}}^2G+\int F^2|\nabla_{j}^\parallel G|^2+
2\int FG(\nabla_{j}^\parallel F)\cdot(\nabla_{j}^\parallel G)
\eea
where $\Delta^\parallel_{j}$ and $\nabla_{j}^\parallel$ are, respectively,  the two
dimensional Laplace operator and gradient.
The term $-\int
F^2G\,\partial_{z_{j}}^2G$ together with $\int V^\perp_{h} F^2G^2$ gives
the confinement energy,
$(e^\perp/h^2) \Vert\Psi\Vert^2$.

Next we consider the first and the third term in \eqref{SCHNEEupp1}.
Since $0\leq f\leq 1$, $f'\geq 0$ and $s_{h}$ is normalized, we have
\begin{align}\notag
\sum_{j}\int F^2|\nabla^\parallel_{j}G|^2+\sum_{j}\int
|\nabla_{j}F|^2 G^2\hfill \nonumber  \leq&
\sum_{j}\int |\nabla_{j}^\parallel \Phi|^2+
 2 \sum_{i<j} f'(|\x_i-\x_j|)^2G^2\ \notag   \\&+
4 \sum_{k<i<j} \int f'(|\x_k-\x_i|)f'(|\x_k-\x_j|) G^2 \label{SCHNEEfpp}
\end{align}
where we have denoted $\prod_{i<j}\varphi(|x_{i}-x_{j}|)$ by $\Phi$
for short. Moreover, since $0\leq \varphi\leq1$, 
\be 2 \sum_{i<j} f'(|\x_i-\x_j|)^2G^2\leq 2 \sum_{i<j} f'(|\x_i-\x_j|)^2s_{h}(z_{i})^2s_{j}(z_{j})^2.\ee

By Young's inequality
\be
2\int_{\R^2} f'(|\x_{i}-\x_{j}|)^2s_{h}(z_{i})^2s_{j}(z_{j})^2
dz_{i}dz_{j}\leq \frac{2\Vert s\Vert^4_4}h\int_{\R} f'(|(x_{i}-x_{j},z)|)^2 dz.
\ee
The right side  gives rise to the first of the two terms in
the formula \eqref{SCHNEEwpot} for the two dimensional
potential $W$. The other term is provided by  $\int
F^2G^2 v_{a}(\x_{i}-\x_{j})$, 
using  $0\leq f\leq1$, $0\leq \varphi\leq1$  and Young's inequality.

Altogether we obtain
\be\label{SCHNEE412}\langle \Psi|H|\Psi\rangle-(ne^\perp/h^2)\langle \Psi|\Psi\rangle
\leq \sum_{j}\int_{\Lambda^n_{2}}
|\nabla_{j}^\parallel \Phi|^2+
\sum_{i<j}\int_{\Lambda^n_{2}}  W(x_{i}-x_{j})\Phi^2 +\mathcal R_{1}+\mathcal R_{2}
\ee
with
\be
\mathcal R_{1}=2\int_{\Lambda^n} FG(\nabla_{j}^\parallel F)
\cdot(\nabla_{j}^\parallel G)
\ee
and
\begin{align}\label{err2}
\mathcal R_{2}&= 4 \sum_{k<i<j} \int_{\Lambda^n} f'(|\x_k-\x_i|)f'(|\x_k-\x_j|)
G^2\\ &\leq \frac 23n(n-1)(n-2)\ell^{2(n-3)}\!\int_{\Lambda^3} \!
f'(|\x_1-\x_2|)f'(|\x_2-\x_3|)s_{h}(z_{1})^2
s_{h}(z_{2})^2s_{h}(z_{3})^2, \notag
\end{align}
where $0\leq \varphi\leq 1$ has been used for the last inequality.

The error term $\mathcal R_{1}$ is easily dealt with: It is zero
because $\varphi(r)$ is constant for $r\leq R$ and $f(r)$ is constant
for $r\geq R$.

The other term, $\mathcal R_{2}$, is estimated as follows. Since
$f'(r)=0$ for $r\geq R$ we can use the Cauchy Schwarz inequality for
the integration over $\x_{1}$ at fixed $\x_{2}$ to obtain
\begin{align}\notag
&\int f'(|\x_{1}-\x_{2}|)s_{h}(z_{1})^2d\x_{1} \\  \notag &\leq
\left(\int f'(|\x_{1}-\x_{2}|)^2
d\x_{1}\right)^{1/2}\left(\int_{|\x_{1}-\x_{2}|\leq R}
s_{h}(z_{1})^4d\x_{1}\right)^{1/2} \\ &\leq (4\pi \Vert s\Vert_{\infty} a'R^3/3h^2)^{1/2}\label{SCHNEE415}
\end{align}
 with
$a'=a(1-a/R)^{-1}$.  The same estimate for the integration over
$\x_{3}$ and a subsequent integration over $\x_{2}$ gives \be \mathcal
R_{2}\leq {\rm (const.)}\ell^{2n} n^3\frac{a'R^3}{\ell^4 h^2}.
\ee
We
need $\mathcal R_{2}/\langle\Psi|\Psi\rangle$ to be small compared to
the leading term in the energy, $\sim n^2\ell^{-2}|\ln(\rho
h^2)|^{-1}$ with $\rho=n/\ell^2$.  (Recall that we are in Region II
where $|\ln(\rho h^2)|\gtrsim h/a$.)  Moreover, the leading term
should be large compared to the Dirichlet localization energy, which
is $\sim n/\ell^2$.  We are thus lead to the conditions (the first
comes from \eqref{SCHNEEnormbd}): \be\label{SCHNEEr1cond}\frac {n^2b^2}{\ell^2}\ll
1,\quad \frac {na'R^3 |\ln(\rho h^2)|} {\ell^2 h^2}\ll 1,\quad
\frac{n}{|\ln(\rho h^2)|}\gg 1, \ee which can also be written
\be\label{SCHNEEr1cond'}{\rho^2\ell^2 b^2}\ll 1,\quad \frac {\rho a'R^3
|\ln(\rho h^2)|} {h^2}\ll 1,\quad \frac{|\ln(\rho
h^2)|}{\rho\ell^2}\ll 1.  \ee These conditions are fulfilled if we
choose \be \label{SCHNEEb} R=h, \quad b=\rho^{-1/2}|\ln(\rho h^2)|^{-\alpha}
\ee with $\alpha>1/2$ and \be\label{SCHNEE419}\rho^{-1/2} |\ln(\rho
h^2)|^{1/2}\ll \ell \ll \rho^{-1/2} |\ln(\rho h^2)|^{\alpha}.\ee Note
also that $n=\rho\ell^2\gg 1$.

It remains to compare
\be\label{SCHNEEphien} \langle\Psi|\Psi\rangle^{-1}\left(\sum_{j}\int_{\R^{2n}}
|\nabla_{j}^\parallel \Phi|^2+
\sum_{i<j}\int_{\R^{2n}}  W(x_{i}-x_{j})\Phi^2\right) \ee
with the expected leading term of the energy, i.e.,
$4\pi (n^2/\ell^2) |\ln(n\atd^2/\ell^2)|^{-1}$.

We consider first the simplest case,
i.e., $n=2$. We have
\be\label{phikin} \int_{\R^2}|\nabla^\parallel \varphi|^2=
(\ln(b/\atd))^{-2}2\pi\int_{R}^b\frac
{dr}r=(\ln(b/\atd))^{-2}2\pi\ln(b/R),\ee
\be\label{phipot}\half\int_{\R^2}W\varphi^2=\frac{4\pi a\Vert s \Vert_4^4}
h\left(\frac{\ln(R/\atd)}
{\ln(b/\atd)}\right)^2.\ee
Inserting the formula \eqref{SCHNEEatd} for $\atd$ and using
$R=h$, $b=\rho^{-1/2}|\ln(\rho h^2)|^{-\alpha}$ and
$a'=a(1+o(1))$ we have
\begin{multline}\label{SCHNEE423}
\int_{\R^2}(|\nabla^\parallel \varphi|^2+\half
W\varphi^2)=\\
2\pi(\ln(b/\atd))^{-2}\left[\ln(b/h)+(h/2a'\|s\|_{4}^4)
\right]=
4\pi |\ln(\rho\atd^2)|^{-1}(1+o(1)).
\end{multline}
For $n>2$ we can use the symmetry of $\Phi$ to write, using \eqref{SCHNEE423}
as well as $0\leq\varphi(r)\leq 1$,
\bea\label{SCHNEE426}&& \sum_{j}\int_{\Lambda_{2}^n}
|\nabla_{j}^\parallel \Phi|^2+
\sum_{i<j}\int_{\Lambda_{2}^n}  W(x_{i}-x_{j})\Phi^2\nonumber\\&&=n
\left(\int_{\Lambda_{2}^{n}}
|\nabla_{1}^\parallel \Phi|^2+
\half\sum_{i=2}^n\int_{\Lambda_{2}^n}  W(x_{i}-x_{1})\Phi^2\right)\nonumber\\
&&\leq 4\pi n^2 \ell^{2(n-1)}
|\ln(n\atd^2/\ell^2)|^{-1}(1+o(1))+\mathcal R_{3}\eea
with
\be \mathcal R_{3}=n^3\ell^{2(n-3)}\int_{\Lambda_{2}^3}
\varphi'(|x_{2}-x_{1}|)
\varphi'(|x_{3}-x_{1}|).\ee
We estimate  $\mathcal R_{3}$
in the same way as
\eqref{SCHNEE415}, obtaining
\be
\mathcal R_{3}\leq \hbox{\rm (const.)}\ell^{2(n-2)}
n^3 b^2 (\ln(b/\atd))^{-2}2\pi\ln(b/R).
\ee
The condition that $\mathcal R_{3}$ has to be
much smaller than the leading term, given by $4\pi n^2 \ell^{2(n-1)}
|\ln(n\atd^2/\ell^2)|^{-1}$, is equivalent to
\be
\frac {n b^2}{\ell^2}\ln(b/R)\ll 1.
\ee
With the choice \eqref{SCHNEEb} this holds if $\alpha>1/2$.
%%%%%%%%%%%%%%%%%

\subsubsection{Upper bound in Region I}

In Region I the ansatz \eqref{SCHNEEFG} can still be
used, but this time we take $b=R$, i.e., $\varphi\equiv 1$.
In this region  $(a/h)|\ln(\rho h^2)|=o(1)$ and the leading term
in the energy is $\sim n^2\ell^{-2}a/h$.
Conditions
\eqref{SCHNEEr1cond} are now replaced by
\be\label{SCHNEEr2cond}
\frac  {n^2R^2}{\ell^2}\ll 1,\quad \frac {nR^3}{\ell^2 h}\ll
1,\quad  \frac{na}{h}\gg 1
\ee
where have here used that $a'=a(1+o(1))$, provided $R\gg a$. Note that
the last condition in \eqref{SCHNEEr2cond} means in particular that $n\gg 1$.
Putting again $\rho=n/\ell^2$, \eqref{SCHNEEr2cond} can  be written as
\be\label{SCHNEEr3cond}
 \rho^2\ell^2R^2\ll 1,\quad \frac {\rho R^3}{h}\ll
1,\quad  \frac{h}{\rho \ell^2 a}\ll 1.
\ee
By assumption, $a/h\ll 1$, but also $\rho a h\ll 1$ by the condition
of strong confinement, c.f. \eqref{SCHNEEahll}. We take
\be\label{SCHNEER1}
R=a(\rho a h)^{-\beta}
\ee
with $0<\beta$, so $R\gg a$. Further restrictions come from the conditions
\eqref{SCHNEEr3cond}: The first and the last of these conditions imply
together that
\be\label{SCHNEEell0}
\frac h{a}\ll \rho\ell^2\ll \frac 1{\rho R^2}
\ee
which can be fulfilled if
\be\label{SCHNEEell1}
\rho a h\ll (\rho a h)^{2\beta}.
\ee
i.e., if $\beta<\half$. Note that this implies in particular $R\ll
\rho^{-1/2}$. We can then take
\be\label{ellupp}
\ell=\rho^{-1/2}(h/a)^{1/2}(\rho a h)^{-\gamma}
\ee
with
\be
0<\gamma<\frac{1-2\beta}2.
\ee

The second of the
conditions \eqref{SCHNEEr3cond} requires that
\be
\frac{\rho R^3}h=(\rho ah)(a/h)^2(\rho a h)^{-3\beta}\ll 1,
\ee
which holds in any case if $\beta\leq 1/3$. A possible choice satisfying all
conditions is
\be
\beta=\frac 13,\quad \gamma=\frac 1{12}.
\ee
The error terms \eqref{SCHNEEr3cond} are then bounded by $(\rho a
h)^{1/6}$ (first and third term) and $(a/h)^2$ (second term).

Finally, with $\Phi\equiv 1$, Eqs. \eqref{SCHNEE412}, \eqref{SCHNEEnormbd} and
\eqref{SCHNEEintw} give
\be \langle\Psi|\Psi\rangle^{-1}
\langle \Psi|H|\Psi\rangle-(ne^\perp/h^2)\leq \frac{4\pi
n^2}{\ell^2}\frac {a\|s\|_{4}^4}h(1+o(1)).
\ee
This completes the proof of the upper bound in boxes with finite $n$.

\subsection{Global bound, uniform case}

The upper bound for the energy per particle in the 2D
thermodynamic limit needed for Theorem \ref{SCHNEEthm:main2D} is
now obtained by dividing $\R^2$ into Dirichlet boxes with side
length $\ell$ satisfying (\ref{SCHNEE419}) in Region II, or
(\ref{SCHNEEell0}) in Region I, and distributing the particles
evenly among the boxes.  In other words, the trial wave function
in a large box of side length $L$ is
\beq\Psi=\sum_{\alpha}\Psi_{\alpha}\eeq where $\alpha$ labels the
boxes of side length $\ell$ contained in the large box, and
$\Psi_{\alpha}$ is the Dirichlet ground state wave function for
$n=\rho\ell^2$ particles in the box $\alpha$, with $\rho=N/L^2$.
The
choice of $\ell$ guarantees in particular that the error
associated with the Dirichlet localization is negligible.  In
order to avoid contributions from the interaction between
particles in different boxes the boxes should be separated by the
range, $R_{0}$ of the interaction potential and in the
``corridors'' between the boxes the wave function is put equal to
zero.  The number of particles in each box is then not
exactly $\rho\ell^2$, but rather the smallest integer larger than
$\rho \ell^2(\ell/(\ell+R_{0}))^2$.  In order that the errors are negligible one needs $R_{0}/\ell=o(1)$, as well as $\rho\ell^2\gg 1$, and both are garanteed by the choice \eqref{ellupp} of $\ell$.

\subsection{Global bound in a trap}

In Theorem \ref{SCHNEEthm:main} the system is inhomogeneous in 2D
because of the trapping potential $V_{L}(x)$, and the distribution
of the particles among the boxes has to be adjusted to the density
given by the GP minimizer $\varphi^{\rm GP}(x)$.  As long as
$\ell$, given by \eqref{ellupp} or \eqref{SCHNEE419}, is small
compared to $L$ this can be done in an analogous way as in
\cite{LSY}, Section 4.2.  Since, however, the condition $\ell\ll
L$ requires ${Na}/h\gg (\bar\rho a h)^{-1/12}$ in Region I, and
$N/{|\ln(\bar \rho h^2)|}\gg |\ln(\bar \rho h^2)|^\varepsilon$ in
Region II, this will not work in the whole parameter range.

A better and only slightly more complicated choice that works in all
cases is to replace the Jastrow-type function
$\prod_{i<j}f(|\x_i-\x_j|)$ by a Dyson wave function of the type
considered in \cite{dyson} and \cite{LSY1999}, i.e., \beq
\label{dysonwave} F(\x_1,\dots,\x_N)=\prod_{i=1}^Nf(t_i(\x_1,\dots,
\x_i))\eeq where \beq t_i(\x_1,\dots, \x_i)=\min_{1\leq j\leq
i-1}|\x_i-\x_j|\eeq is the distance between $\x_i$ and its nearest
neighbor among the points $\x_1,\dots,\x_{i-1}$.  To take care of the
inhomogeneity in the 2D variables a factor $\prod_i\varphi^{\rm
GP}(x_i)$ is included, where $\varphi^{\rm GP}$ the normalized
minimizer of the 2D GP functional with coupling constant $Ng$.
Moreover, the local behavior in the 2D variables is modelled by a
Dyson wave function \beq
\Phi(x_{1},\ldots,x_{N})=\varphi(t'_{i}(x_{1},\ldots,x_{N}))\eeq with
\beq t'_i(x_1,\dots, x_i)=\min_{1\leq j\leq i-1}|x_i-x_j|.\eeq The
global trial wave function is thus of the form \beq
\label{uppansatz1}\Psi(\x_1,\dots,\x_N)=F(\x_1,\dots,\x_N)G(\x_1,\dots,\x_N)\eeq
with $F$ given by \eqref{dysonwave}, and \beq \label{uppansatz2}
G(\x_1,\dots,\x_N)=\Phi(x_{1},\ldots,x_{N})\prod_i \varphi^{\rm
GP}(x_i) s_h(z_i).\eeq The advantage of the Dyson wave functions over
the Jastrow-type functions is that when computing the expectation
values $\langle\Psi,H\Psi\rangle/\langle\Psi,\Psi\rangle$
cancellations between numerator and denominator take place that
effectively improve the estimate \eqref{SCHNEEnormbd} of the norm of
$\Psi$.

\subsubsection{Region I}
In Region I, where one can take $\varphi\equiv 1$, the computation of
the expectation value of the Hamiltonian is the same as in
\cite{LSY1999}.  In fact, there is even no need to do the computations
explicitly because the required bound can be obtained by combining
Theorem 2.1 with Theorem III.1 in \cite{LSY1999}.  One checks that the
parameter $b$ in Eq.\ (3.42) in \cite{LSY1999} is proportional to
$\bar\rho^{-1/3}h^{1/3}$ and hence the error terms in the upper bound
of $E^{\rm QM}$ in terms of the 3D GP energy are of the order $a/b=
(a/h)^{2/3}(\bar\rho ah)^{1/3}$.  On the other hand, by Theorem 2.1
the 3D GP energy minus confining energy approaches, as $h/L\to 0$, the
2D GP energy with coupling constant $(\int s^4)a/h$ and the
approximation is uniform in the parameters as long as $(\bar\rho ah)\to
0$.  Altogether we have the bound \beq \label{uppregionI}E^{\rm
QM}(N,L,h,a)-Nh^{-2}e^{\perp}\leq E^{\rm GP}_{\rm 2D}(N,L,(\hbox
{$\int s^4$})a/h)(1+o(1))\eeq where the error term o(1) tend to zero
if $h/L\to 0$, $a/h\to 0$ and $\bar\rho ah\to 0$.

\subsubsection{Region II}
In Region II one uses the ansatz
\eqref{uppansatz1}--\eqref{uppansatz2} with $f$ as in \eqref{eff} with
$R=h$ and $\varphi$ as in \eqref{varphi} with
$b=\bar\rho^{-1/2}|\ln(\bar\rho h^2)|^\alpha$.  Since the computations
are analogous to those in \cite{LSY1999} and in Eqs.\ (3.11)-(3.29)
above we shall not write all details explicitly.  Proceeding
as in Section 3.1 in \cite{LSY1999}, using \eqref{SCHNEEupp1}, we
obtain for the expectation value of the Hamiltonian
\eqref{SCHNEEdef:ham}
\begin{equation}\label{exp}
\begin{split}
\frac{\langle\Psi,H^{(N)}\Psi\rangle}{\langle\Psi|\Psi\rangle}&\leq
2\suli_{i=1}^N\frac{\int |\Psi|^2F_i^{-2}f'(t_i)^2}{\int
|\Psi|^2}+\suli_{j<i}\frac{\int |\Psi|^2v(\xij)}{ \int |\Psi|^2}\\
&+2\suli_{k\leq
i<j}\frac{\int
|\Psi|^2|\eps_{ik}\eps_{jk}|F_i^{-1}F_j^{-1}f'(t_i)f'(t_j)}{\int
|\Psi|^2}\\ &+\suli_{i=1}^N\frac{\int F^2G(-\nabla^2_i
G)+|\Psi|^2\V_{L,h}(\x_i)}{\int |\Psi|^2}.
\end{split}
\end{equation}
where
\begin{equation}
\eps_{ik} =\begin{cases}
1 &\text{for $i=k$}\\
-1 &\text{for $t_i=|x_i-x_k|$}\\
0 &\text{otherwise}
\end{cases}.
\end{equation}
The next step is to decouple a given pair of variables, $\x_{i}$,
$\x_{j}$, from the other variables in order to achieve cancellations
between numerator and denominator in the terms on the right side of
\eqref{exp}. As in \cite{LSY1999} one defines for $i<j<p$
\beq F_{p,i}=f(\min_{k<p,k\neq i} |\x_p-\x_k|), \qquad
F_{p,ij}=f(\min_{k<p,k\notin\{i,j\}} |\x_p-\x_k|).\eeq
Note that $F_{p,i}$ is independent of $i$ and $F_{p,ij}$ is
independent of $i$ and $j$. As in Eqs.\ (3.13)--(3.15) in
\cite{LSY1999} one has
\begin{align}
&F_{j+1}^2\dots F_{i-1}^2F_{i+1}^2\dots F_N^2\leq F_{j+1,j}^2\dots
F_{i-1,j}^2F_{i+1,ij}^2\dots F_{N,ij}^2 \label{above}\\
\intertext{and}\label{below} &F_j^2\dots F_N^2\geq
F_{j+1,j}^2\dots F_{i-1,j}^2F_{i+1,ij}^2\dots F_{N,ij}^2\\ \nonumber
&\times\left(1-\suli_{k=1,\, k\neq i,j}^N(1- f(|\x_j-\x_k|)^2)
\right)
\left(1-\suli_{k=1,\, k\neq i}^N(1-f(|\x_i-\x_k|)^2)\right).
\end{align}
In an analogous way one defines $\Phi_{p,i}$ and $\Phi_{p,ij}$ and
obtains the corresponding equations \eqref{above} and \eqref{below} with $f$ replaced by
$\varphi$ and $\x$ by the 2D variable $x$.

Consider now the first two terms in \eqref{exp}.  As in \cite{LSY1999}
one uses the estimates \begin{equation} f'(t_i)^2\leq\suli_{j=1}^{i-1}
f'(|\x_i-\x_j|)^2,\quad{\rm and}\quad F_i\leq f(|\x_i-\x_j|).\eeq For fixed
$i,j$ one can use \eqref{above} and \eqref{below} (and their analogues
for $\Phi$) to separate the contribution from the variables $\x_{i}$ and $\x_{j}$
from the rest of the integrand.  The part that depends only on the
other variables cancels between numerator and denominator, while the $ij$
contribution in the numerator is
\begin{multline}\label{wenerg}\int
\left(2f'(|\x_{i}-\x_{j}|)^2+v(|\x_{i}-\x_{j}|)f(|\x_{i}-\x_{j}|)^2\right)
\\ \times
s_{h}(z_{i})^2s_{h}(z_{j})^2\varphi(|(x_{i}-x_{j}|)^2\varphi^{\rm
GP}(x_{i})^2\varphi^{\rm GP}(x_{j})^2d\x_{i}d\x_{j}.\end{multline} Integration
over $z_{i}$ and $z_{j}$, using Young's inequality, bounds
\eqref{wenerg} by \beq\label{wenerg2}\int
W(x_{i}-x_{j})\varphi(|(x_{i}-x_{j}|)\varphi^{\rm
GP}(x_{i})^2\varphi^{\rm GP}(x_{j})^2.\eeq where $W$ is defined in
\eqref{SCHNEEwpot}.  This, in turn, can as in Eqs.\ (3.18)--(3.20) in
\cite{LSY1999} be bounded by \beq 2\int\varphi^{\rm GP}(x)^4dx J\eeq
with \beq J=\half\int W(x)\varphi(|x|)^2dx\eeq given by \eqref{phipot}
The matching kinetic term, $\int |\nabla^{\parallel}\varphi|^2$, given
by \eqref{phikin} ist derived form the last term in \eqref{exp} in the
same way.  These two terms together give, up to small errors, the
correct coupling constant, i.e., the factor before $\int \varphi^{\rm
GP}(x)^4dx$ in the 2D GP functional.  The other parts of the 2D GP functional
as well as the confining energy $Nh^{-2}e^{\perp}$ follow
from the last term in \eqref{exp}.

It remains to look at the error terms.  These come on the one hand
from the $ij$ contribution to the denominator, i.e., the last two
factors in \eqref{below} and the corresponding factors with
$\varphi$ instead of $f$.  On the other hand they come from the
third term in \eqref{exp} and corresponding terms with $\varphi$
instead of $f$. The first mentioned errors have (cf.  Eq.\ (3.21)
in \cite{LSY} and Eq.\ (3.1) in \cite{LSY2d}) the form \beq
N\Vert\varphi^{\rm GP}\Vert_{\infty}^2\Vert s_{h}\Vert_{\infty}^2
(4\pi R^3/3),\quad{\rm and}\quad N\Vert\varphi^{\rm
GP}\Vert_{\infty}^2\pi b^2.\eeq Since $\Vert\varphi^{\rm
GP}\Vert_{\infty}^2\sim \bar L^{-2}$ and $\Vert
s_{h}\Vert_{\infty}^2\sim h^{-1}$, while $R=h$ and
$b=\bar\rho^{-1/2}|\ln(\bar\rho h^2)|^{-\alpha}$, these terms are
of the order $\bar\rho h^2$ and $|\ln(\bar\rho h^2)|^{-2\alpha}$
respectively.  For the other errors one again exploits the
cancellations between numerator and denominator and obtains in the
same way as in \cite{LSY} and \cite{LSY2d} (cf.\ Eqs.  (3.26) and
(3.36) in \cite{LSY1999}, and (3.4)--(3.5) in \cite{LSY2d}) terms
of the order $(\bar\rho a h)^2$, $N^{-1}$, and $|\ln(\bar\rho
h^2)|^{-1}$.  This can be compared with the estimate for $\mathcal
R_{2}$ in \eqref{err2} where cancellations could not be used.  All
the mentioned error terms are small in Region II. In combination
with the bound \eqref{uppregionI} for Region I we can thus state
the upper bound in both regions as \beq \label{uppregionII}E^{\rm
QM}(N,L,h,a)-Nh^{-2}e^{\perp}\leq E^{\rm GP}_{\rm
2D}(N,L,g)(1+o(1)).\eeq

\section{Scattering length}\label{pertscatt}

As a preparation for the lower bound we consider in this section the
perturbative calculation of the 2D scattering length of an integrable
potential.

Consider a 2D, rotationally symmetric potential $W\geq 0$ of finite
range $R_{0}$.  As discussed in Appendix A in \cite{LY2d} the
scattering length is determined by minimizing, for
$R\geq R_{0}$, the functional \be\label{SCHNEEscattfunct}{\mathcal E}_{R}[\psi]=\int_{|x|\leq
R}\left\{|\nabla \psi|^2+\half W|\psi|^2\right\}\ee with boundary
condition $\psi=1$ for $|x|=R$.  The Euler equation (zero energy
scattering equation) is \be\label{SCHNEEscatt}-\Delta \psi+\half W \psi=0\ee
and for
$r=|x|\geq R_{0}$ the minimizer, $\psi_{0}$, is \be\psi_{0}(r)=\ln(r/a_{\rm
scatt})/\ln(R/a_{\rm scatt})\ee with a constant $a_{\rm scatt}$.  This
is, by definition, the 2D scattering length for the potential $W$.  An
equivalent definition follows by computing the energy, \be
E_{R}={\mathcal E}_{R}[\psi_{0}]=2\pi/\ln(R/a_{\rm scatt})\ee which
means that \be\label{SCHNEE3.5} a_{\rm scatt}=R\exp(-2\pi/E_{R}).\ee
\begin{lem}[Scattering length for soft
potentials]\label{SCHNEEscattlemm}
    Assume $W(x)=\lambda w(x)$ with $\lambda \geq 0$, $w\geq 0$, and
    $w\in L^1(\R^2)$, with $\int w(x)dx=1$.  Then, for $R\geq R_{0}$,
    \be \label{SCHNEEscattcorreq1}a_{\rm
    scatt}=R\exp\left(-\frac{4\pi+\eta(\lambda,R)}{\lambda}\right)\ee
    with $\eta(\lambda,R)\to 0$ for $\lambda\to 0$.
    \end{lem}
    \begin{proof}
The statement is, by \eqref{SCHNEE3.5},  equivalent to \be E_{R}=\half
\lambda(1 +o(1))  \ee
where the error term may depend on $R$. The
upper bound is clear by the variational principle, taking $\psi=1$ as
a test function.  For the lower bound note first that $\psi_{0}\leq
1$.  This follows from the variational principle: Since $W\geq 0$ the
function $\tilde\psi_{0}(x)=\min\{1,\psi_{0}\}$ satisfies ${\mathcal
E}_{R}[\tilde\psi_{0}]\leq {\mathcal E}_{R}[\psi_{0}]$.  Hence the
function $\varphi_{0}=1-\psi_{0}$ is nonnegative.  It satisfies
\be\label{SCHNEEphieq}-\Delta \varphi_{0}+\half W \varphi_{0}=\half W\ee and
the Dirichlet boundary condition $\varphi_{0}=0$ for $|x|=R$.

Integration of
 (\ref{SCHNEEscatt}), using that $\psi_{0}(r)=1$ for $r=R$, gives
\be E_{R}=\half\int W\psi_{0}=\half\int
W(1-\varphi_{0}).\ee
Since $\varphi_{0}\geq 0$ we thus need to show
that $\|\varphi_{0}\|_{\infty}=o(1)$.

By \eqref{SCHNEEphieq}, and since $\varphi_{0}$ and $W$ are both
nonnegative, we have $-\Delta\varphi_{0}\leq\half W$ and hence
\be \varphi_{0}(x)\leq
\int K_{0}(x,x')W(x')dx'\ee where $K_{0}(x,x')$ is the (nonnegative)
integral kernel
of
$(-\Delta)^{-1}$ with Dirichlet boundary conditions at $|x|=R$.  The
kernel $K_{0}(x,x')$ is integrable (the singularity is
$\sim \ln |x-x'|$) and hence, if $W$ is bounded, we have
$\|\varphi_{0}\|_{\infty}\leq {\rm(const.)}\lambda
\|w\|_{\infty}=O(\lambda)$.

If $w$ is not bounded we can, for every $\varepsilon>0$, find a bounded
$w^\varepsilon\leq w$ with $\int (w-w^\varepsilon)\leq \varepsilon$.
Define $C_{\varepsilon}=\|w^\varepsilon\|_{\infty}$.  Without
restriction we can assume that $C_{\varepsilon}$ is a monotonously
decreasing function of  $\varepsilon$ and continuous.  The function
$g(\varepsilon)=\varepsilon/C_{\varepsilon}$ is then monotonously
increasing in $\varepsilon$ (and hence decreasing if $\varepsilon$ decreases), continuous and $g(\varepsilon)\to 0$ if
$\varepsilon\to 0$.  For every (sufficiently small) $\lambda$ there is
an $\varepsilon(\lambda)=o(1)$ such that $g(\varepsilon(\lambda))=\lambda$.
Then \be\|\varphi_{0}\|_{\infty}\leq
{\rm(const.)}(\varepsilon(\lambda) +\lambda
C_{\varepsilon(\lambda)})={\rm(const.)}\varepsilon(\lambda)=o(1).\ee
\end{proof}

\begin{corollary}[Scattering length for
scaled, soft
potentials]\label{SCHNEEscattcorr}
    Assume $W_{R,\lambda}(x)=\lambda R^{-2} w_{1}(x/R)$ with
    $w_{1}\geq 0$ fixed and $\int w_{1}=1$. Then the scattering length
    of $W_{R,\lambda}$ is
    \be \label{scattcorreq} a_{\rm
    scatt}=R\exp\left(-\frac{4\pi+\eta (\lambda)}{\lambda}\right)\ee
    with $\eta (\lambda)\to 0$ for $\lambda\to 0$, independent of $R$.
    \end{corollary}
    \begin{proof} This follows from Lemma \ref{SCHNEEscattlemm}
    noting that, by scaling, the scattering length of $W_{R,\lambda}$ is $R$ times
    the scattering length of $\lambda w_{1}$. The latter is
    independent of~$R$.
\end{proof}

\noindent{\it Remark.} If $W$ is obtained by averaging a 3D
integrable potential $v$ over an interval of length $h$ in the $z$
variable, the formula (\ref{scattcorreq}), together with
Eq.\ (A.8) in \cite{LY2d}, motivates the exponential dependence of
the effective 2D scattering length
(\ref{SCHNEEatd}) of $v$ on $h/a$: The integral $\lambda=\int W(x)dx$ is
$h^{-1}\int
v(\x)d^3\x$, which for weak potentials is $h^{-1}8\pi a$ to lowest
order, by Eq.\ (A.8) in \cite{LY2d}.
Inserting this into (\ref{scattcorreq}) gives (\ref{SCHNEEatd})
(apart from the dependence on the shape function $s$). This heuristics is, of course,
only valid for soft potentials $v$. An essential step in the
lower bound in the next section is the replacement of $v$ by a soft
potential to which this reasoning can indeed be applied.

\section{Lower bound}

\subsection{Finite boxes}

Like for the upper bound we
consider first
the homogeneous case and finite boxes, this time with Neumann
boundary conditions.  The optimal distribution of particles among the
boxes is determined by using sub\-additivity and convexity arguments
as in \cite{LY1998} and \cite{LSY}.

In the treatment of the lower bound there is a natural division
line between the case where the mean particle distance
$\rho^{-1/2}$ is comparable to or larger than $h$ and the case
where it is much smaller than $h$.  The first case includes Region
II and a part (but not all) of Region I. When $\rho^{-1/2}$ is
much smaller than $h$ the boxes have finite extension in the $z$
direction as well. The method is  then a fairly simple
modification of the 3D estimates in \cite{LSY1999} (see also
Section 4.4 in \cite{LSY}) and will not be discussed further here.

The
derivation of a lower bound for the case that $\rho h^2\leq
C<\infty$ proceeds by the following steps:
\begin{itemize}
    \item
     Use Dyson's Lemma \cite{dyson, LY1998} to replace $v_a$ by an integrable 3D potential
$U$, retaining part of the kinetic energy.

\item Average the potential $U$ at fixed $x\in\R^2$ over the
$z$-variable to obtain a 2D potential $W$.  Estimate the error
in this averaging procedure by
using Temple's inequality \cite{TE, RS} at each fixed $x$.

\item The result is a 2D many body problem with an integrable
interaction potential $W$
which, by Corollary \ref{SCHNEEscattcorr},
has the right 2D scattering length to lowest order in $a/h$,
but reduced kinetic energy inside the range of the potential.
 This problem is treated in the same way as in
\cite{LY2d, LSY2d}, introducing a 2D Dyson potential and using perturbation
theory, again estimating the errors by
Temple's inequality.

\item Choose the parameters (size $\ell$ of box, fraction
$\varepsilon$ of the kinetic energy, range $R$ of potential $U$, as
well as the corresponding parameters for the 2D Dyson potential)
optimally to minimize the errors.
\end{itemize}

The first two steps are analogous to the corresponding
steps in the proof of the lower bound in Theorem 3.1 in \cite{LSY}, cf.\
Eqs.\ (3.30)--(3.36) in \cite{LSY}. It is, however, convenient to define
the Dyson potential $U$ in a slightly different manner than in
\cite{LY1998}. Namely, for $R\geq 2R_{0}$, with $R_{0}$ the range
of $v_a$, we define
\begin{equation}\label{softened2}
U_{R}(r)=\begin{cases}\frac {24}7 R^{-3}&\text{for
$\half R<r<R$ }\\
0&\text{otherwise.}
\end{cases}
\end{equation}
The reason is that this potential has a simple scaling with $R$ which
is convenient when applying Corollary \ref{SCHNEEscattcorr}.
Proceeding  as in Eqs.\ (3.30)--(3.38) in \cite{LSY}  we write a
general
wave function as
\beq\label{loweransatz}
\Psi(\x_1,\dots,\x_n)=f(\x_1,\dots,\x_n)\prod_{k=1}^n
s_h(z_k) \ ,
\eeq
and define $F(x_1,\dots,x_n)\geq 0$ by
\beq\label{defF2}
|F(x_1,\dots,x_n)|^2=\int |f(\x_1,\dots,\x_n)|^2\prod_{k=1}^n
s_h(z_{k})^2 dz_k\ .
\eeq
Note that $F$ is normalized if $\Psi$ is normalized.
The analogue of Eq.\ (3.38) in \cite{LSY} is
\begin{eqnarray}\nonumber
&&\langle\Psi|H|\Psi\rangle- \frac{n e^\perp}{h^2}\geq \\ \nonumber
&& \sum_{i=1}^n \int\Big[\eps |\nabla^\parallel_i F|^2 +
(1-\eps)|\nabla^\parallel_i F|^2 \chi_{\min_k |x_i-x_k|\geq R}(x_i)
\Big]\prod_{k=1}^n dx_k \\ \nonumber &&+\sum_{i=1}^n \int \left[
\eps|\partial_i f|^2 + a'
U_{R}(|\x_i-\x_{k(i)}|)\chi_{{\mathcal B}_\delta}(z_{k(i)}/h)|f|^2
\right] \prod_{k=1}^n s_h(z_k)^2 d \x_k , \\ \label{57a}
\end{eqnarray}
where  $\nabla^\parallel_{i}$ denotes the gradient with respect to
$x_{i}$
and $\partial_j=d/dz_j$. Moreover,
$\chi_{{\mathcal B}_\delta}$ is the characteristic function of the subset
 ${\mathcal B}_\delta\subset \R$
where
$s(z)^2\geq \delta$ for $\delta>0$,
\beq \label{aprime}
a'=a(1-\eps)(1-2 R \|\partial s^2\|_\infty/
(h \delta)),\eeq and $k(i)$ denotes the index of the nearest neighbor to $\x_{i}$.
When deriving \eqref{57a} the Cauchy Schwarz inequality has been used to bound the longitudinal kinetic energy of $f$ in terms of that of $F$, i.e.,
\be\label{cauchy}
\sum_{i=1}^n \int|\nabla^\parallel_i f|^2\prod_{k=1}^n s_h(z_k)^2 d \x_k \geq
\sum_{i=1}^n \int|\nabla^\parallel_i F|^2\prod_{k=1}^n  d x_k.\ee

We now consider, for fixed $x_1,\dots,x_n$, the term
\beq\label{44a}
\sum_{i=1}^n \int \left[
\eps|\partial_i f|^2 + a'
U_{R}(|\x_i-\x_{k(i)}|)\chi_{{\mathcal B}_\delta}(z_{k(i)}/h)|f|^2
\right] \prod_{k=1}^n s_h(z_k)^2 d z_k \ .
\eeq
This can be estimated from below by the expectation value of
$U_R$ over $z_i$ at fixed $x_i$, using Temple's inequality to estimate the errors.
The result is, by a
calculation analogous to Eqs.\ (3.41)--(3.46) in \cite{LSY},
\begin{eqnarray}\nonumber
\langle\Psi|H|\Psi\rangle- \frac{ne^\perp}{h^2} &\geq& \int\sum_{i=1}^n
\Big[\eps |\nabla^\parallel_i F|^2 +(1-\eps)|\nabla^\parallel_i F|^2 \chi_{
\min_{k\neq i}|x_i-x_k|\geq R}(x_i)\nonumber
 \\
\label{SCHNEE:putt} & & +\half \sum_{j\neq i}W(x_{i}-x_{j})|F|^2 \Big]\prod_{k=1}^n dx_k\,,
\end{eqnarray}
where $W$ is obtained by
averaging $a'U_{R}$ over $z$:
\beq \label{w} W(x-x')=2a^{\prime\prime}\int_{\R\times\R} s_h(z)^2 s_h(z')^2 U_{R}(|\x-\x'|)
\chi_{{\mathcal B}_\delta}(z'/h) dz dz' \ .
\eeq
Here, $a^{\prime\prime}=a'(1-\eta)$ with an error term $\eta$
containing the error estimates from the Temple inequality and 
from ignoring other points than the nearest neighbor to $\x_{i}$. Moreover,
since $\int U_{R}(\x)d\x=4\pi$, $U_{R}(|\x-\x'|)=0$ for $|\x-\x'|>R$, and
$|s(z)^2-s(z')^2|\leq R \Vert \partial_{z}s^2\Vert_{\infty}$ for
$|z-z'|\leq R$ we have the
estimate
\beqa \int_{\R^2} W(x) dx &\geq &
\frac{8\pi a^{\prime\prime}}
h \left(\int_{\mathcal B_{\delta}}s(z)^4dz-\frac Rh
\Vert \partial_{z}s^2\Vert_{\infty}\right)\nonumber\\
&\geq& \frac{8\pi a^{\prime\prime}}
h\left(\Vert s\Vert^4_{4}-\delta-\frac Rh
\Vert \partial_{z}s^2\Vert_{\infty}\right).\label{Wlower}
\eeqa

As
we will explain in a moment, the
errors, and the replacement of $n-1$ by $n$, require the following
terms to be small:
\begin{equation}\label{SCHNEEerr:3d}\frac{nh^2a}{\varepsilon R^3},\quad
    \frac{nR}{ h}, \quad
\varepsilon, \quad
 \frac1n,  \quad \delta, \quad \frac R{h\delta}.
\end{equation}
%%%%%%%%%%%%%
The rationale behind the first term is as follows. The Temple errors in
the averaging procedure at fixed $\x_{1},\ldots,x_{n}$ produces a
factor similar to \eqref{SCHNEEtemplegp}, namely
\beq\label{60}
\left(1- a'\frac{\langle
U^2\rangle}{\langle U\rangle}\frac{1}{({\rm const.})\varepsilon/h^2-({\rm const.})
a'\langle U\rangle}\right)
\eeq
with
\beq
\langle U^m\rangle=  \int \left(\sum_{i=1}^n
U(|\x_i-\x_{k(i)}|)\chi_{{\mathcal B}_\delta}(\x^\perp_{k(i)}/r)\right)^m
\prod_{j=1}^n s_h(z_{j})^2 dz_{j},
\eeq
cf.\ Eqs. (3.40)--(3.41) in \cite{LSY}. The analogue of Eq. (3.42)
in \cite{LSY} is
\be
\langle U\rangle \leq ({\rm const.})
n(n-1)\frac{\|s\|_4^4}{h R^2},
\ee
and $\langle
U^2\rangle\leq ({\rm const.}) nR^{-3}\langle U\rangle$ by Schwarz's
inequality. Since the denominator in \eqref{60} must be positive, we
see in particular that the particle number must satisfy
\be \label{nbound1} n(n-1)<({\rm const.})\frac{\varepsilon R^2}{ah}\ee
and the error is of the order ${nh^2a}/{\varepsilon R^3}$ as claimed
in \eqref{SCHNEEerr:3d}.

The estimate from below on $\langle U\rangle$, obtained in the same
way as Eq.\ (3.46) in \cite{LSY}, is
\beqa\nonumber
\langle U\rangle\!\!
&\geq& \!\! \sum_{i\neq j} \int
U(|\x_i-\x_j|)\chi_{\mathcal B_\delta}(z_j/h)\left(1-\!\!\! \sum_{k,
\, k\neq i,j} \theta(R-|\x_k-\x_i|)\right)\prod_{l=1}^n
s_{h}(z_l)^2 dz_{l} \\ &\geq& \!\!
\frac1{2a^{\prime\prime}}\sum_{i\neq j}W(x_i-x_j) \left(1-(n-2)
\frac{R}{h} \|s\|_\infty^2\right) \ .
\eeqa
In particular, the second term in \eqref{SCHNEEerr:3d}, $nR/h$, should be small.
The requirement that $\varepsilon$ and $n^{-1}$ are small
needs no further comments.

The potential $W$ can be written as
\beq\label{Wpot}
W(x)=\lambda R^{-2}w_{1}(x/R)\,,
\eeq
where $w_{1}$ is independent of $R$, with \beq \int w_{1}(x)dx=1\eeq
and \beq\label{lambda}\lambda=\frac{8\pi\hbox{ $a\int
s^4$}}h(1-\eta')\,.\eeq
Here, $\eta'$ is an error term involving $\delta$ and
$R/(h\delta)$ (cf.\ (\ref{aprime}) and \eqref{Wlower}) besides the
other terms in
(\ref{SCHNEEerr:3d}). In particular, $\delta$ and $R/(h\delta)$
should be small. The 2D
scattering length of (\ref{Wpot}) can be computed by
Corollary \ref{SCHNEEscattcorr} and has the right form (\ref{SCHNEEatd})
to leading order in $\lambda$. (Recall from the remark preceding Eq.\
(\ref{ggprime}) that $R$ in (\ref{scattcorreq}) can be replaced by $h$
as long as $ca<R<Ch$.)

The Hamiltonian on the right side of Eq.\ (\ref{SCHNEE:putt}) can
now be treated with the methods of \cite{LY2d}.  The only
difference from the Hamiltonian discussed in that paper is the
reduced kinetic energy inside the range of the potential $W$. This
implies that $\lambda$ in the error term $\eta(\lambda)$ in Corr.\
\ref{SCHNEEscattcorr} should be replaced by $\lambda/\varepsilon$,
which requires \be
\frac a{\varepsilon h}\ll 1.\label{ahe}\ee
Otherwise the method is the same as in \cite{LY2d}: a slight
modification of Dyson's Lemma (lemma 3.1 in \cite{LY2d}) allows to
substitute for $W$ a potential $\tilde U$ of larger range, $\tilde
R$, to which perturbation theory and Temple's inequality can be
applied.  The modified Dyson lemma is discussed in the Appendix.
The fraction of the kinetic energy borrowed for the application of
Temple's inequality as in Eq.\ (3.16) in \cite{LY2d} will be
denoted by $\tilde \varepsilon$. The errors that have now to be
controlled are
\begin{equation}\label{err:2d}
\tilde\varepsilon,   \quad n\tilde R^2/\ell^2, \quad \frac{R}{\tilde R},
\quad \frac{n\ell^2}{\tilde\varepsilon
\tilde R^2\ln(\tilde R^2/\atd^2)} .
\end{equation}

To explain these terms we refer to Eqs.\ (3.18)--(3.19) in \cite{LY2d}
which contain the relevant estimates.  Substituting $\tilde
\varepsilon$ for $\varepsilon$ and $\tilde R$ for $R$ in these
inequalities we see first that $\tilde \varepsilon$ and $n\tilde
R^2/\ell^2$ should be small.  
The smallness of $R/\tilde R$ guarantees that the \lq\lq hole" of radius $R$ in the 2D Dyson potential has negligible effect.
Since the denominator in the Temple error in
Eq.\ (3.19) in \cite{LY2d} must be positive we see also that the particle
number $n$ in the box should obey the bound \be n(n-1)<({\rm
const.})\,\tilde\varepsilon\ln(\tilde R^2/\atd^2)\label{nbound2},\ee
and the Temple error is bounded by $({\rm const.})
{n\ell^2}/({\tilde\varepsilon \tilde R^2\ln(\tilde R^2/\atd^2)})$.

We
summarize the discussion so far in the following Lemma:

\begin{lem} For all $n$ satisfying \eqref{nbound1} and \eqref{nbound2}
    \be E_{\rm box}(n,\ell,h,a)\geq \frac{2\pi n(n-1)}{\ell^2}|\ln (a_{{\rm
    2D}}^2/\tilde R^2)|^{-1}\left(1-\mathcal E(n,\ell,h,a;
    \varepsilon,R,\tilde\varepsilon,\tilde R,\delta)\right)\label{boxlower}\ee
    where $\mathcal E$ tends to zero together with terms listed in
    \eqref{SCHNEEerr:3d}
    and \eqref{err:2d}.
    \end{lem}

    We note also from Eq.\
(3.19) in \cite{LY2d} that $K(n)=(1-\mathcal E)$  is decreasing in $n$ (for
    the other parameters fixed). If $K(n)$ is defined as zero for
    $n$ not satisfying \eqref{nbound1} and
    \eqref{nbound2} the estimate \eqref{boxlower} hold for all $n$.

\subsection{Global bound, uniform case}

Using superadditivity of the energy (which follows from $W\geq
0$), monotonicity of $K(n)$ and convexity of $n(n-1)$ in the same
way as in Eqs.\ (7)--(12) in \cite{LY1998},  one sees that if
$\rho=N/L^2$ is the density in the thermodynamically large box of
side length $L$ the optimal choice of $n$ in the box of fixed side
length $\ell$ is $n\sim \rho \ell^2$.  We thus have to show that
it is possible to choose the parameters $\varepsilon$, $R$,
$\delta$, $\tilde\varepsilon$ and $\tilde R$ and $\ell$,in such a
way that all the error terms (\ref{SCHNEEerr:3d}) and
(\ref{err:2d}), as well as $\delta$ and $R/(h\delta)$ are small,
while $|\ln (a_{{\rm
    2D}}^2/\tilde R^2)|=|\ln (a_{{\rm
    2D}}^2\rho)| (1+o(1))$.
We note that the conditions
$a/h\ll 1$ and $\rho|{\ln(\rho\atd^2)}|^{-1}\ll 1/h^2$ imply
$\rho\atd^2\to 0$ and hence $|\ln(\rho\atd^2)|^{-1}\to 0$.

We make the ansatz
\be \label{2dparam}\varepsilon=\left(\frac ah\right)^\alpha, \quad \delta=
\left(\frac ah\right)^{\alpha'}, \quad
R=h\left(\frac
ah\right)^\beta,
\ee
and choose $\ell$ such that $L$ is a multiple of $\ell$ with
\be \rho^{-1/2}\ll \ell\lesssim\rho^{-1/2}\left(\frac
ah\right)^{-\gamma}.\label{ell}\ee
Then $n=\rho\ell^2\gg 1$ and
$R/(h\delta)=(a/h)^{\beta-\alpha'}$. The error terms (\ref{SCHNEEerr:3d}) are also
powers of $a/h$ and we have to ensure that all
exponents are positive, in particular
\be\label{errexponents} \beta-\alpha'>0,
\quad\beta-2\gamma>0,\quad{1-\alpha-3\beta-2\gamma}>0.
\ee
This is fulfilled, e.g., for
\be \label{2dexpo}\alpha=\alpha'=\frac19,\quad \beta=\frac29,\quad\gamma=\frac1{18}\,,\ee
with all the exponents (\ref{errexponents}) equal to 1/9. Note also
that with this choice Eq.\ \eqref{ahe} is fulfilled.

Next we write \be \tilde\varepsilon=|\ln(\rho\atd^2)|^{-\kappa},
\quad \tilde R=\rho^{-1/2}|\ln(\rho\atd^2)|^{-\zeta}. \ee Then
$|\ln (a_{{\rm
    2D}}^2/\tilde R^2)|=|\ln (a_{{\rm
    2D}}^2\rho)| (1+o(1))$.  The error terms \eqref{err:2d} are
\be\label{{errest:2d}}
\tilde\varepsilon=|\ln(\rho\atd^2)|^{-\kappa}, \quad \frac{R}{\tilde R}
=\left(\frac ah\right)^\beta (\rho h^2)^{1/2}|\ln(\rho\atd^2)|^{\zeta},
\quad
\frac{n\tilde R^2}{\ell^2}=
|\ln(\rho\atd^2)|^{-2\zeta},
\ee and
\be\label{2dtemperr} \frac{n\ell^2}{\tilde\varepsilon
\tilde R^2\ln(\tilde R^2/\atd^2)}=
\left(\frac ah\right)^{-4\gamma}|
\ln(\rho\atd^2)|^{-(1-\kappa-2\zeta)}(1+O(\ln|\ln(\rho\atd^2)|)).
\ee
Since $\left(a/h\right)^{-4\gamma}|
\ln(\rho\atd^2)|^{-4\gamma}=O(1)$, the error term \eqref{2dtemperr} can also
be written
as
\be\frac{n\ell^2}{\tilde\varepsilon
\tilde R^2\ln(\tilde R^2/\atd^2)}=
O(1)|\ln(\rho\atd^2)|^{-(1-\kappa-2\zeta-4\gamma)}(1+O(\ln|\ln(\rho\atd^2)|)).
\ee
The condition $\rho h^2<C$ is used to bound $R/\tilde R$ in \eqref{{errest:2d}}.
Namely,  \be(\rho h^2)^{1/2}|\ln(\rho\atd^2)|^{\zeta}\leq {\rm
(const.)}(h/a)^{\zeta},\ee
so
\be
\frac{R}{\tilde R}
=O(1)\left(\frac ah\right)^{\beta-\zeta}.
\ee
We choose now
\be  \zeta=\frac 19,\quad\kappa=\frac 29.\ee
Then
\be
\beta-\zeta=\frac 19\quad\text{and}\quad 1-\kappa-2\zeta-4\gamma=\frac 13.
\ee

This completes our discussion of the lower bound of the energy in
the thermodynamic limit, i.e., in Theorem \ref{SCHNEEthm:main2D},
for the case  $\rho h^2\leq C<\infty$. As already mentioned, the
case $\rho h^2\gg 1$ can be treated with the 3D methods of
\cite{LY1998, LSY1999}, taking care to retain uniformity in the
parameter $h/\ell$. The analogous problem for the reduction form
3D to 1D is discussed in detail in \cite{LSY}, cf.\ in particular
the lower bound in Theorem 3.2 in \cite{LSY}. Since no new aspects
arise in the reduction from 3D to 2D we refrain from discussing
the case $\rho h^2\gg 1$ further here.

\subsection{Global bound in a trap and BEC}

We now discuss the case when the system is inhomogeneous in the $x$
variables, i.e., the lower bound for Theorem \ref{SCHNEEthm:main}.  A
possible approach, analogous to that of \cite{LSY1999}, is to modify
Eq.\ \eqref{loweransatz} and write a general wave function as
\beq\label{loweransatz2}
\Psi(\x_1,\dots,\x_n)=f(\x_1,\dots,\x_n)\prod_{k=1}^n\varphi^{\rm
GP}(x_{k}) s_h(z_k) \ , \eeq This is always possible because
$\varphi^{\rm GP}$ and $s_{h}$ are strictly positive.  As in
\cite{LSY1999} the task then becomes to minimize a quadratic form in
$f$ involving the GP density $\rho^{\rm GP}(x_{i})=\varphi^{\rm
GP}(x_{i})^2$ both as a weight factor in the integration over the
$x_{i}$'s and also as a replacement for the external potential
$V_{L}(x_{i})$.  There is, however, a neat alternative approach to deal
with inhomogeneities due to R. Seiringer \cite{norway} that is
somewhat simpler and that we shall follow with appropriate
modifications.  This approach leads also quickly to a proof of Theorem
\ref{BEC}.

\subsubsection{GP case}

We start by writing again the wave function in the form
\eqref{loweransatz} but
this time with $n=N$. With $H$ including the trapping potential
$V_{L}$, cf. \eqref{SCHNEEdef:ham}, and using the symmetry of $F$
and $f$
we obtain as in
Eq.\ \eqref{57a}
\begin{eqnarray}\nonumber
\langle\Psi|H|\Psi\rangle- \frac{N e^\perp}{h^2}\hskip -.5cm&&\geq T+
N \int V_{L}(x_{1})F^2\prod_{k=1}^N dx_k
\\ \nonumber &&+N \int \left[
\eps|\partial_1 f|^2 + a'
U_{R}(|\x_1-\x_{k(1)}|)\chi_{{\mathcal B}_\delta}(z_{k(1)}/h)|f|^2
\right] \prod_{k=1}^N s_h(z_k)^2 d \x_k , \\ \label{57b}
\end{eqnarray}
where
\be T=N \left[\eps \int_{|x_{1}-x_{k(1)}|\leq R}|\nabla^\parallel_1 \Psi|^2
\prod_{k=1}^N d\x_k
+\int_{|x_{1}-x_{k(1)}|\geq R}
|\nabla^\parallel_1 \Psi|^2 \prod_{k=1}^N d\x_k
\right].\ee
Note that we have here not yet used Eq. \eqref{cauchy} to bound the
parallel kinetic energy of $f$ in terms of that of $F$. The reason is that we want to prove BEC for the wave function $\Psi$ and not only for $F$.

The next step is to split the kinetic energy $T$,
for given $\tilde R>R$ and $\tilde\varepsilon >0$,
into two parts with the main contributions coming from length scales $>\tilde
R$ and $<\tilde R$ respectively in the longitudinal directions.
To write this in a compact
way we introduce the notation
\be { X}=(x_{2},\ldots,x_{N}), \quad d{X}=dx_{2}\cdots
dx_{N}; \quad Z=(z_1,...,z_N),\quad dZ=dz_1\cdots dz_N\ee
and
\begin{multline} \Omega_{\X}^{<R}=\{x_{1}\,:\,|x_{1}-x_{k(1)}|\leq
R \}, \quad \Omega_{\X}^{R,\tilde R}=\{x_{1}\,:\,R\leq
|x_{1}-x_{k(1)}|\leq
\tilde R \}, \\ \Omega_{\X}^{>\tilde R}=\{x_{1}\,:\,
|x_{1}-x_{k(1)}|\geq
\tilde R \}.\end{multline}
The splitting of the kinetic energy is
\be T\geq T^{>}+T^{<}\label{kinsplitting}\ee
with
\beq
T^{>}= N\int_{{\mathbb
R}^{3N-2}}\left[\frac{\varepsilon\tilde\varepsilon}2\int_{\Omega_{\X}^{<R}}
|\nabla_{1}^{\parallel}\Psi|^2 {}
+\frac{\tilde\varepsilon}2\int_{\Omega_{\X}^{R,\tilde R}}
|\nabla_{1}^{\parallel}\Psi|^2 {}
+\big (1-\frac{\tilde\varepsilon}{2}\big)\int_{\Omega_{\X}^{>\tilde R}}
|\nabla_{1}^{\parallel}\Psi|^2 {}\right]d\X dZ,\\
\eeq
\beq
T^{<}= N\int_{{\mathbb
R}^{2(N-1)}}\left[\varepsilon\big(1-\frac{\tilde\varepsilon}{2}\big)\int_{\Omega_{\X}^{<R}}
|\nabla_{1}^{\parallel}F|^2 {}
+\big(1-\frac{\tilde\varepsilon}{2}\big)\int_{\Omega_{\X}^{R,\tilde R}}
|\nabla_{1}^{\parallel}F|^2 {}
+ \int_{\Omega_{\X}^{>\tilde R}}\frac{\tilde\varepsilon}2
|\nabla_{1}^{\parallel}F|^2 {}\right]d\X.\\
\eeq Here we have used \eqref{cauchy} for $T^<$, but not for
$T^>$. We then add and subtract $8\pi N^2g\int_{{\mathbb
R}^{2N}}\rho^{\rm GP}(x_{1})|F|^2$ from the right side of
\eqref{57b} and, using \eqref{kinsplitting}, estimate it by
$E^{>}+E^{<}$ with \be\label{egross} E^{>}=T^{>}+N \int
V_{L}(x_{1})F^2+8\pi N^2g\int\rho^{\rm GP}(x_{1})|F|^2, \ee
\begin{multline}\label{eklein}
E^{<}=T^{<}+N\int \left[
\eps|\partial_1 f|^2 + a'
U_{R}(|\x_1-\x_{k(1)}|)\chi_{{\mathcal B}_\delta}(z_{k(1)}/h)|f|^2
\right] \prod_{k=1}^N s_h(z_k)^2 d \x_k\\-8\pi N^2g\int\rho^{\rm GP}(x_{1})|F|^2.
\end{multline}
The two terms are now considered separately. The first remark is
that $\varphi^{\rm GP}(x)$ is the ground state wave function of
the 1-body Hamiltonian \be-\Delta^{\parallel}+V_{L}(x)+8\pi
Ng\rho^{\rm GP}(x)\ee and the energy is equal to the GP chemical
potential \be\label{gpchem} \mu^{\rm GP}(Ng)=E^{\rm GP}_{\rm
2D}(1,L,Ng)+4\pi Ng\int\varphi^{\rm GP}(x)^4 dx,\ee cf. Section 2
in \cite{LSY2d}. Hence, if $T^{>}$ were replaced by the full
kinetic energy, $\int |\nabla_{1}^{\parallel}\Psi|^2$, then
$E^{>}$ would be bounded below by $N\mu^{\rm GP}(Ng)=E^{\rm
GP}_{\rm 2D}(N,L,g)+4\pi N^2g\int\varphi^{\rm GP}(x)^4 dx$. Using
a variant of the generalized Poincar\'e inequalities of
\cite{poincare} it is shown in \cite{rot2, norway} that this
estimate also holds for $E^{>}$ up to small errors. Even a
stronger result is true (cf.\ \cite{norway}, Eq.\ (51)), because a
contribution form the one particle density matrix in the
lontitudinal direction, \be \gamma^{\parallel}_{\Psi}(x,x')
%= \int_{\R^{3N-4}}\Psi((x,z),(\X,Z))
%\Psi((x',z),(\X,Z))^*d\X dZ
=\int_{\R}\gamma_\Psi((x,z),(x',z))dz\ee
can be included:
Provided $\tilde R^2 N\to 0$ for $N\to\infty$, and $Ng$ is fixed,
\be
E^>/N\geq \left(\mu^{\rm GP}(Ng)+C{\rm
Tr}\left[\mbox{$\frac{1}{N}$}\gamma^{\parallel}_{\Psi}(1-|\varphi^{\rm
GP}\rangle\langle\varphi^{\rm
GP}|)\right]\right)(1-o(1))\label{elarge}
\ee
where the term $o(1)$ goes to zero if $\varepsilon,\tilde\varepsilon\to
0$ go to zero after the limit $N\to\infty$
has been taken.

The term $E^<$ is treated in the same manner as in the homogeneous
case by introducing boxes, $\Lambda_\alpha$, of side length $\ell$
in the longitudinal directions. Defining
\be\rho_\alpha=\sup_{x\in\Lambda_\alpha}\rho^{\rm GP}(x) \ee we
have (cf.\ Eq.\ (60) in \cite{norway}) \be
\label{eestim}E^<\geq \inf_{n_\alpha}\sum_\alpha
(E(n_\alpha,\ell)-8\pi Ng \rho_\alpha n_\alpha) \ee with $E(n,\ell)$
the right side of \eqref{boxlower}. (It depends also on the
parameters $h,a;\epsilon,R,\varepsilon,\tilde R, \delta$.) The
infimum is taken over all distributions of the $N$ particles among
the boxes and $n_\alpha$ denotes the particle number in box
$\alpha$. Choosing the box length as in Eq. \eqref{ell} with
$\rho$ replaced by the mean density $\bar\rho\sim N/\bar L$, and
arguing exactly as in \cite{norway}, Eqs.\ (71)--(78) we obtain
\be E^<\geq -4\pi Ng\int|\varphi^{\rm
GP}(x)|^4dx(1+o(1)).\label{esmall} \ee Adding \eqref{esmall} and
\eqref{elarge} we obtain the bound \be\label{lowerinhomo}
\frac{\langle\Psi|H|\Psi\rangle}N- \frac{e^\perp}{h^2} \geq
\left[E^{\rm GP}_{\rm 2D}(1,L,Ng)+C{\rm
Tr}[\hbox{$\frac1N$}\gamma^{\parallel}_{\Psi}(1-|\varphi^{\rm
GP}\rangle\langle\varphi^{\rm GP}|)\right](1-o(1)). \ee If $\Psi$
is the ground state wave function $\Psi_{0}$, then
\eqref{lowerinhomo} together with the upper bound
\eqref{uppregionII} proves both the convergence of the energy as
claimed in Theorem \ref{SCHNEEthm:main} and also the convergence
of the density matrix $\gamma^{\parallel}_{\Psi}(x,x')$ to
$\varphi^{\rm GP}(x)\varphi^{\rm GP}(x')$ (in trace norm)  as
$N\to \infty$ for the case that $Ng$ is fixed (or at least
bounded). The convergence of the full density matrix
$\gamma_{\Psi_{0}}(\x,\x')$, scaled by $h$ in the variables $z$
and $z'$, to $\varphi^{\rm GP}(x)\varphi^{\rm GP}(x') s(z)s(z')$
follows from a simple energetic consideration: The energy gap
between the lowest and the first excited level of
$-\partial_{z}^2+V^\perp_h$ is large compared to
${\langle\Psi|H|\Psi\rangle}/N- \frac{e^\perp}{h^2}$ in the limit
considered. Hence the contributions to the density matrix form the
excited states in the perpendicular direction must vanish.

\textit{Remark.} The above proof of BEC can be extended to allow a
slow increase of $Ng$ to infinity with $N$, but this improvement
is restricted by the available estimates of the error terms in
\eqref{elarge} and \eqref{lowerinhomo} and is only marginal.

\subsubsection{TF case}
To complete the proof of  Theorem \ref{SCHNEEthm:main} one must
also consider the case that $Ng\to\infty$, i.e. the `Thomas-Fermi'
limit. Here the starting point is again \eqref{57b}, this time
with $T$ replaced by the smaller quantity \be T'=\left[\eps
\int_{|x_{1}-x_{k(1)}|\leq R}|\nabla^\parallel_i F|^2
\prod_{k=1}^N dx_k +\int_{|x_{1}-x_{k(1)}|\geq R}
|\nabla^\parallel_i F|^2 \prod_{k=1}^N dx_k \right]\ee cf.\
\eqref{cauchy}. We then use the TF equation \be \rho^{\rm
TF}(x)=\frac1{8\pi Ng}[\mu^{\rm TF}-V_{L}(x)]_{+}\label{TF}\ee to
trade the potential $V_{L}(x)$ for the TF density $\rho^{\rm
TF}(x)$ which is the square of the minimizer of
\eqref{SCHNEE2dgp}  with the kinetic term omitted. The TF chemical
potential $\mu^{\rm TF}$ is determined by the normalization
condition $\int \rho^{\rm TF}=1$ and satisfies in analogy to
\eqref{gpchem} \be\label{tfchem}\mu^{\rm TF}(Ng)=E^{\rm TF}_{\rm
2D}(1,L,Ng)+4\pi Ng\int\varphi^{\rm TF}(x)^4 dx,\ee From
\eqref{TF} it follows that \be V_{L}(x)\geq \mu^{\rm TF}-8\pi (Ng)
\rho^{\rm TF}(x).\ee Both $\mu^{\rm TF}$ and $ \rho^{\rm TF}$
depend on $Ng$ and scale in a simple way if the potential
$V_{L}(x)$ is a homogeneous function of some degree $p>0$, cf.
Eqs.\ (2.20)-(2.21) in \cite{LSY}. We can now apply the box method
to the minimization of
\begin{multline} \label{energytf}T'+\mu^{\rm TF}-8\pi (Ng)  N\int \rho^{\rm TF}(x_{1})
F^2\prod_{k=1}^N dx_k\\+N \int \left[
\eps|\partial_1 f|^2 + a'
U_{R}(|\x_1-\x_{k(1)}|)\chi_{{\mathcal B}_\delta}(z_{k(1)}/h)|f|^2
\right] \prod_{k=1}^N s_h(z_k)^2 d \x_k,\end{multline}
obtaining in analogy to \eqref{eestim} and using \eqref{tfchem},
\be
(\ref{energytf})\geq E^{\rm TF}_{\rm 2D}(N,L,g)+4\pi (Ng)N\int\rho^{\rm
TF}(x)^2 dx+
\inf_{n_\alpha}\sum_\alpha (E(n_\alpha,\ell)-8\pi Ng \rho^{\rm TF}_\alpha
n_\alpha).
\ee
The remaining steps towards the estimate
\be
(\ref{energytf})\geq E^{\rm TF}_{\rm 2D}(N,L,g)(1-o(1))
\ee
are now exactly like in Section 4 in \cite{LSY2d}.

\section{Conclusions} We have investigated the dimensional reduction
of a trapped, interacting  Bose gas when the trap potential is
strongly confining in one direction. Starting from the many-body
Hamiltonian with repulsive, short range
interactions we have shown rigorously how an effective 2D description
of the ground state energy and density
emerges and how the parameters of the 2D gas relate to those of the
3D gas. Two parameter regimes can be distinguished: One where the
gas retains some of its 3D character despite the tight
confinement and another where the situation is manifestly
two-dimensional with
a logarithmic dependence of the coupling parameter on the
density. Moreover, we have shown that the trapped gas is Bose-Einstein
condensed in the ground state provided the coupling parameter $Ng$ stays
bounded.

\appendix
\section{Appendix: A modification of Dyson's lemma}

Let $W(x)\geq 0$ be a rotationally symmetric potential in 2D with
$W(x)=0$ for $|x|>R$.  Denote by $B_{R}$  the ball of radius $R$ around the origin.
 For $0<\varepsilon$ and $R'\geq R$ define
 \be E_{R',\varepsilon}=\min \int_{B_{R}}
\left[\varepsilon |\nabla \phi(x)|^2 +
\half W(x)|\phi(x)|^2\right]dx+\int_{B_{R'}\setminus B_{R}}
|\nabla \phi(x)|^2 dx
\label{E1}
\ee
where the minimum is taken over $\phi\in H^1(B_{ R'})$
with $\phi(x)=1$ for $|x|= R'$.

\begin{lem}
  \be\label{A.2}
  E_{R',\varepsilon}=\frac{2\pi}{\ln(R'/R)+2\pi/E_{R,\varepsilon}}
  \ee
    \end{lem}
    \begin{proof} For any $c>0$ the minimum of the first integral in \eqref{E1} with boundary condition $\phi(x)=c$ for $|x|=R$ is $c^2E_{R,\varepsilon}$ and the minimum of the second integral with the same boundary condition at $|x|=R$ and $\phi(x)=1$ for $|x|= R'$ is $2\pi\ln(R'/R)/(\ln(\tilde R/b))^2$ where $b$ is determined by $\ln (R/b)/\ln(R'/b)=c$. Adding the two contributions and minimizing over $c$ gives \eqref{A.2}.\end{proof}

\begin{lem}[Modified Dyson Lemma]\label{moddysonl}
    Let $W$ be as above, let $\tilde R>R$
    and let ${\tilde U}(r)\geq 0$ be any function with support in $[R,\tilde R]$
    satisfying
\begin{equation}\label{1dyson}
 2\pi\int_{R}^{\tilde R} {\tilde U}(R')E_{R',\varepsilon}^{-1}R'dR'\leq 1
\end{equation}
with $E_{R',\varepsilon}$ as in \eqref{A.2}.
Let ${\mathcal B}\subset \R^2$ be star-shaped  with respect
to $0$.
Then, for all functions $\phi  \in H^1(\mathcal{B})$,
\begin{multline}
\int_{\mathcal B\cap B_{R}}
\left[\varepsilon |\nabla \phi(x)|^2 +
\frac{1}{2}W(x)|\phi(x)|^2\right]dx+\int_{\mathcal B\cap
(B_{\tilde R}\setminus B_{R})}
|\nabla \phi(x)|^2 dx \\
\geq   \int_{\cal B} {\tilde U}(|x|)
|\phi (x)|^2 ~dx.
\label{2dyson}
\end{multline}
\end{lem}

\begin{proof}
The proof is very similar to that of Lemma 3.1 in \cite{LY2d}. In
polar coordinates, $r,\theta$, one has
$|\nabla \phi|^2 \geq |\partial \phi /\partial r|^2$ so it
suffices to
prove the analogue of \eqref{2dyson} for each angle $\theta \in [0,2\pi)$.
Denote
$\phi (r,\theta)$  simply by $f(r)$, and let
$R(\theta)$ denote the distance of the origin to the boundary
of $\mathcal{B}$ along the ray $\theta$. It suffices to consider
the case $R\leq R(\theta)$ (here,
$W\geq0$ is used) and the estimate to prove is
\begin{multline} \label{radial}
\int_0^{R} \left\{\varepsilon |\partial f(r) /\partial r|^2 +
\half W(r)|f(r)|^2\right\} ~rdr +
\int_{R}^{\min\{\tilde R,R(\theta)\}} |\partial f(r) /\partial r|^2r\,
dr\\ \geq
 \int_{R}^{\min\{\tilde R,R(\theta)\}}  {\tilde U}(r)|f(r)|^2 ~rdr.
\end{multline}
For the given
value of $\theta$, consider the disc $B_{R(\theta)}$ centered at the
origin in $\mathbb{R}^2$ and of radius $R(\theta) $.
Our function $f$ defines a rotationally
symmetric function, $x\mapsto f(\vert x\vert)$ on
$B_{R(\theta)}$, and \eqref{radial} is
equivalent to
\begin{multline}\label{disc}
\int_{B_{R(\theta)}\cap B_{{R}_{0}}} \left[\varepsilon|\nabla f
(|x|)|^2 + \frac{1}{2}W(r)|f(|x|)|^2\right]dx+
\int_{B_{R(\theta)}\cap (B_{\tilde R}\setminus B_{{R}_{0}})}|\nabla f
(|x|)|^2 dx\\
\geq \int_{B_{R(\theta)}} {\tilde U}(|x|)|f(|x|)|^2 dx
\end{multline}

If $R\leq R'\leq {\min\{\tilde R,R(\theta)\}}$ the left side of
\eqref{disc} is not smaller than the same quantity with
$B_{R(\theta)}$ replaced by the smaller disc $B_{R'}$.
(Again, $W\geq 0$ is used.)
According to \eqref{E1}
this integral over  $B_{R'}$ is at least
$E_{R',\varepsilon}|f(R')|^2$. Hence, for every such $R'$,
\begin{multline} \label{pointwise}
\int_0^{R} \left\{\varepsilon |\partial f(r) /\partial r|^2 +
\half W(r)|f(r)|^2\right\} ~rdr +
\int_{R}^{\min\{\tilde R,R(\theta)\}} |\partial f(r) /\partial r|^2r\,
dr\\ \geq
(2\pi)^{-1}
E_{R',\varepsilon}|f(R')|^2
 \end{multline}
The proof is completed by multiplying both sides of
\eqref{pointwise} by $2\pi {\tilde
U}(R')E_{R',\varepsilon}^{-1}R'$ and, finally, integrating with
respect to $R'$ from $R$ to ${\min\{\tilde R,R(\theta)\}}$.
\end{proof}

A convenient choice for the Dyson potential ${\tilde U}$ is
\beq\label{defU} {\tilde U}(r) = \left\{ \begin{array}{ll} \nu(\tilde
R)^{-1} & {\rm if\ }R\leq r\leq \tilde R \\ 0 & {\rm otherwise} \ ,
\end{array}\right.  \eeq with \be \nu(\tilde R)=2\pi\int_{R}^{\tilde
R}E_{R',\varepsilon}R'\,dR'.\ee 

In the case considered in this paper the potential $W$ is integrable with coupling constant $\lambda \sim a/h(1+o(1))$ while $\varepsilon\sim (a/h)^{1/9}$. In particular $\lambda/\varepsilon\to 0$ as $a/h\to 0$. From  Section 4 and the discussion of Eqs.\  \eqref{lambda}--\eqref{ahe}
 we thus  conclude that in our case
\be E_{R,\varepsilon}=2\pi/\ln(R/a_{\rm 2D})(1+o(1)).\ee
By
Eq.\  \eqref{A.2} we then also have
\be E_{R',\varepsilon}=2\pi/\ln(R'/a_{\rm 2D})(1+o(1)).\ee
and thus  \be \nu(\tilde
R)={\mbox{$\frac{1}{4}$}}\tilde R^2\ln(\tilde
R^2/a_{\rm 2D}^2)(1+o(1)).\label{nuR}\ee
Thus, for  $R$ and $\varepsilon$ chosen as in \eqref{2dparam} and \eqref{2dexpo}, the modified Dyson Lemma gives, up to negligible errors,  exactly the same result as the standard 2D Dyson Lemma for a potential with scattering length $a_{\rm 2D}$.\\

\noindent{\bf Acknowledgements.} We thank Robert Seiringer and Elliott Lieb for helpful
discussions.  KS thanks the Institute for Theoretical Physics, ETH
Z\"urich for hospitality and JY the Science Institute of the
University of Iceland, Reykjavik, where parts of this work were done. 
This work is supported by the Post Doctoral Training Network
HPRN-CT-2002-00277 of the European Union and a grant P17176-N02 of the
Austrian Science Fund (FWF).

\end{document}